\documentclass[1p]{elsarticle}

\journal{Journal of \LaTeX\ Templates}

\usepackage{amsmath,amssymb,amsthm,amsfonts,afterpage}
\usepackage{amscd,subeqnarray}
\usepackage{array}
\usepackage[export]{adjustbox}

\usepackage{boldfonts}
\usepackage{booktabs}
\usepackage{color}

\usepackage{graphics}
\usepackage{lineno}
\usepackage{multirow}
\usepackage{natbib}
\usepackage{ragged2e}

\usepackage{soul}
\usepackage{subfig}
\usepackage{stmaryrd}
\usepackage{symbols}

\usepackage{tikz, xcolor}
\usetikzlibrary{shapes.geometric}
\usetikzlibrary{shapes,arrows}

\usepackage{wrapfig} 
\usepackage{xcolor,cancel}

\newcommand{\bfa}[1]{\boldsymbol{#1}} 			
			
\newcommand{\bfeps}{\boldsymbol{\epsilon}}

\newcommand{\bfsig}{\boldsymbol{\sigma}}

\newcommand{\curl}{\text{curl}}   				%
\newcommand{\tr}{\text{tr}}       				%

\DeclareMathAlphabet{\mathpzc}{OT1}{pzc}{m}{it}

\newcommand{\bfx}{\boldsymbol{x}}

\newtheorem{theorem}{Theorem}[section]

\newtheorem{remark}[theorem]{Remark}

\begin{document}

\begin{frontmatter}

\title{Quasi-Static Anti-Plane Shear Crack Propagation in a New Class of Nonlinear Strain-Limiting Elastic Solids using Phase-Field Regularization}

\author[TAMU]{Hyun C. Yoon}
\textcolor{cyan}{\ead{hyun.yoon@tamucc.edu}}

\author[FL]{Sanghyun Lee}
\ead{lee@math.fsu.edu}

\author[TAMU]{S. M. Mallikarjunaiah\corref{cor1}}
\ead{M.Muddamallappa@tamucc.edu}

\cortext[cor1]{Corresponding author}
\address[TAMU]{
Department of Mathematics \& Statistics,
Texas A\&M University-Corpus Christi
6300 Ocean Drive, Unit 5825
Corpus Christi, TX 78412-5825}
\address[FL]{
Department of Mathematics,
Florida State University, 
208 Love Building
1017 Academic Way
Tallahassee, FL 32306}

\begin{abstract}
We present a novel {constitutive} model using {the} 
framework of strain-limiting theor{ies of elasticity} for an evolution of quasi-static anti-plane fracture. 
The classical linear elastic fracture mechanics (LEFM), with conventional linear relationship between stress and strain,
has {a well documented inconsistency through which it predicts a singular crack-tip strain}. {This clearly violates the basic tenant of the theory which is a first order approximation to finite elasticity.}  
To overcome the issue, we investigate a new class of material models which predicts uniform and bounded strain throughout the body.
The nonlinear model allows the strain value to remain small even if the stress value tends to infinity,
which is achieved by {an} implicit relationship between  stress and strain. 
{A major objective of this paper is to couple a nonlinear bulk energy with diffusive crack 
employing the phase-field approach.} 
{Towards that end, an iterative L-scheme is employed and the numerical model is augmented with a  penalization technique to accommodate irreversibility of crack.} 
Several numerical experiments are presented to illustrate the capability and the performance of the proposed framework   
We observe the naturally bounded strain {in the neighborhood} of the crack-tip, leading to different bulk and crack energies for fracture propagation. 
\end{abstract}


\end{frontmatter}

\section{Introduction}

{Brittle} crack or fracture in structures has been drawing a great amount of attention from various fields of research from civil to mechanical, even coupled with electrical engineering, as it may bring severe impacts to the structure it evolves upon. 
In terms of functionality, environments and safety in every civilized society, such damage or collapse can be caused particularly due to the property of brittleness of fracture. On the other hand, certain engineers are devoted to actually foster the brittle fracture to utilize it for an industrial purpose, such as the recent hydraulic fracturing in petroleum industry \cite{Bourdin2018,hubbert1972mechanics,IPACS_2020}. Thus, accurately identifying crack growth mechanism is very important in every design application. 

A vast amount of studies have been devoted to the brittle fracture propagation models. First and foremost, there is the linear elastic fracture mechanics (LEFM). 
This celebrated model can also be classified as the Griffith-Irwin approach~\cite{broberg1999}, since it was by Griffith's idea that gave a birth of LEFM to engineering community \cite{griffith1921phenomena,ValP1996}. Although Griffith did not pour much attention to the area of crack-tip, the brilliant concept of energy differentiated the fracture energy from the bulk energy for an existing crack to propagate \cite{broberg1999,griffith1921phenomena}. Further, the stress intensity factor was introduced by Irwin as the criterion for the crack growth: if it reaches the critical stress intensity factor or the fracture toughness, the static crack of a slit or a shorter dent  transforms to grow \cite{IrwG1957}. Thus,  for a brittle elastic material where any dissipation is only from the bulk energy, LEFM approximates the growth of existing crack based on the linear relation between stress and strain.

However, it is true that the
kinematics involved at the crack-tip may be located far beyond the area of classical elasticity. Some reports reveal experimental evidences about the nonlinear behavior of non-dissipative materials such as titanium alloys  \cite{zhang2009fatigue} even within the small strain regime. For LEFM's calculation of the Cauchy stress and the linearized strain under the assumption of small strain, the classical Hooke's law (or linear elastic constitutive law) is employed with the linear approximation to the general theory of elasticity \cite{OgdR1984}. 
Since the relationship or constitutive law is based on Hooke's law, the strain cannot avoid being calculated as proportional to the singular behavior of Cauchy stress. Accordingly, it results in some unrealistic values and additionally all the nonlinear behavior of cracks (e.g., the coalescence) cannot be explained accurately. Not to mention the inaccuracy for the crack growth model, this unbounded strain near the crack-tip itself contradicts the assumption of small strain theory for a deformable body within the scale of continuum mechanics.

To overcome the issue and accommodate the experimental reality near the crack-tip, various models are investigated and proposed. One approach is to manipulate the tip area along with its scale, such as by introducing  two dimensional cohesive zones or three dimensional process zones in the vicinity of the strain concentrators as crack-tips \cite{broberg1999}. Then, the phenomena are reduced to dislocations with the scale down even to the atoms, where an autonomy is established for the cohesive energy and stress \cite{barenblatt1962mathematical}. The strain and stress are calculated in the zone inside of which the linear continuum equations break down into the microscopic size. In fact, LEFM can similarly treat the yielding within the small scale calculations \cite{broberg1999, RicJ1974, BroW2018}. These representations are, however, more based on some ad-hoc treatments with its difficulty to validate experimentally, which has some other drawbacks such that it is not only the partial expressions for the phenomena but the pre-processing procedures are often required.
 
Beyond the classical relations, a novel and broader class of elasticity for the Cauchy and Green formulations has been studied focusing on the relationship between the stress and strain \cite{rajagopal2003implicit, rajagopal2007elasticity}. Then, a nonlinear relationship can be established for the stress and the linearized small strain by introducing the implicit constitutive theories. Utilizing the fundamental approach, recently it has been investigated to model the stress-strain behavior for non-dissipative elastic solids in 
 \cite{rajagopal2011non,rajagopal2011conspectus,rajagopal2014nonlinear,rajagopal2007response}. 
 {More recently, a special subclass of  isotropic nonlinear and non-dissipative models have been studied for a single anti-plane shear crack \cite{rajagopal2011modeling} and a plane-strain crack \cite{gou2015modeling}. The results in both \cite{rajagopal2011modeling}  and \cite{gou2015modeling} indicate that the strains remain bounded at the crack-tip.}

In this study, we aim to utilize the aforementioned nonlinear models to investigate  quasi-static evolution of fracture. 
To this end, we couple the novel material constitutive model with regularized variational mechanics, which is also known as phase-field approach. From the definition of phase-field, the discontinuous interface of a crack is turned into the diffusive zone around a crack. 
One of the main advantages of the phase-field is that no additional constitutive rules or criteria are required that govern when a crack should nucleate, grow, change direction, or merge/split into multiple cracks, but only through the minimization of energy functional.  In particular, computing additional stress intensity factors near the fracture tips is intrinsically embedded in the model. Moreover, the energy functional is based on the classical Griffith's theory and LEFM for brittle fracture \cite{griffith1921phenomena,barenblatt1962mathematical}. In this regard, the phase-field approach for modeling the fracture has received a lot of attention from the applied mechanics community.

Some recent successful relevant phase-field literatures include thermal shocks and thermo-elastic-plastic solids \cite{BouMarMauSics14,MieHofSchaeAl15,noii2019phase}, elastic gelatin for wing crack formation \cite{LeeRebHayWhe_2016}, pressurized fractures \cite{BourChuYo12,MiWheWi15b,MiWheWi14,MiWheWi15c,WiLeeWhe15,Wi16_fsi_pff}, fluid-filled (i.e., hydraulic) fractures \cite{MiWheWi14,LeeWheWi16,Miehe2015186,MieheMauthe2015,Markert2015,heider2016phase},
proppant-filled fractures \cite{LeeMiWheWi16}, variably saturated porous media \cite{Cajuhi2017}, crack initiations with microseismic probability maps \cite{LeeWheWiSri16,wheeler2019unconventional}, and many other applications~\cite{Bourdin2018,choo2018cracking,Yoshioka2019,LeeWheWi16,heider2016phase,mandal2019phase,LeeMiWheWi16,LeeMikWheWick2017_pftwo,LeeMinWhe2018}.

The governing system for the mechanics and the phase-field are developed based on the Euler-Lagrange formulation. Due to the irreversibility constraint from the phase-field energy functional, the augmented Lagrangian method~\cite{fortin2000augmented,glowinski1989augmented,wheeler2014augmented}  is discussed.  Both nonlinear mechanics and nonlinear phase-field equations are linearized through Newton iterations. 
In addition, for the fully coupled system, we employ a recently investigated operator splitting scheme \cite{brun2019iterative},  L-scheme, to decouple the operators for computing efficiency. We note that the phase-field function could be utilized as the indicator function for further adaptive mesh refinement techniques.

We find that the advantage of the nonlinear strain-limiting model with the energy minimization using phase-field over the classical models in that the strain remains small
even if the stress tends to very large values, which is believed to be critical for accurate modeling for fracture propagation. More importantly, satisfying the assumption of small strain theory and the experimental results  \cite{zhang2009fatigue}, the model proposed is logically consistent in its derivation of generic form for a 
nonlinear relationship between the Cauchy stress and linearized strain through the implicit constitutive theory. Several numerical examples comparing the nonlinear strain-limiting model and classical linear elasticity model for  quasi-static fracture in mode-III are presented and utilized to evaluate the performance of the new model. We find different physical responses between the classical LEFM and the proposed nonlinear strain-limiting model. The material behavior focusing on the crack-tip are compared and different fracture propagation with the bulk and the crack energies are obtained. 

The remaining organization of the paper is as follows: In Section~\ref{sec:background}, we briefly introduce the derivation of strain-limiting model  and recapitulate the main idea of phase-field approach, where the mathematical model (governing system) for our problem is discussed.
Spatial and temporal discretization using finite element method and the solution algorithm are presented in Section~\ref{sec:num}.
Finally, several numerical examples comparing the classical LEFM model and nonlinear strain-limiting model for quasi-static fracture propagation are illustrated in Section \ref{sec:examples}.

\section{Mathematical Model} 
\label{sec:background}
In this section, we introduce basic kinematics that are needed for our problem description. Then we develop a modeling framework based on strain-limiting theories elasticity and a variational approach for quasi-static fracture propagation with the phase-field regularization.

\subsection{Kinematics for elasticity and strain-limiting theories }

Let $\mathcal{B}$ be a fixed domain in the reference configuration representing a stress-free  elastic body, with a given boundary $\partial \mathcal{B}$. The boundary is decomposed into the displacement boundary ($\partial \mathcal{B}_D$) and the traction boundary ($\partial \mathcal{B}_N$), which satisfy $\partial \mathcal{B} = \bar{\partial \mathcal{B}_D} \cup \bar{\partial \mathcal{B}_N}$ and $\partial \mathcal{B}_D \cap \partial \mathcal{B}_N = \emptyset$. 
Let $\bx:= f(\bX)$ denote the current (or deformed) position of a particle (motion of a particle) that is at $\bX$ in a stress-free reference configuration $\mathcal{B}$ of a material body.  
Here $f$ is a deformation of the body which is differentiable and the displacement is denoted by 
$\bu := \bx - \bX$.
The displacement gradients are defined as
\begin{equation}
\dfrac{\partial \bu}{\partial \bX} := \nabla_\bX \bu = \bF - \bI 
\ \ 
\text{ and } 
\ \
\dfrac{\partial \bu}{\partial \bx} := \nabla_\bx \bu =  \bI - \bF^{-1}, 
\end{equation}
where $\bI$ is the identity matrix and $\bF$ is the deformation gradient as
\begin{equation}
\bF := \dfrac{\partial f}{\partial \bX}.
\end{equation}
The Cauchy-Green stretch tensors $\bB$ and $\bC$ are given by 
\begin{equation}
\text{(left) } \ \bB := \bF \bF^{\text{T}}, \ \  \text{(right) } \  \bC :=  \bF^{\text{T}} \bF, 
\end{equation}
respectively. 
Then the Green-St.Venant strain tensor $\bE$ and the Almansi-Hamel strain $\be$ are defined as
\begin{equation}
\bE := \dfrac{1}{2} (\bC - \bI) \ \ \text{ and } \ \
\be := \dfrac{1}{2} (\bI - \bB^{-1}). 
\end{equation}
Under the assumption of small displacement gradients such that, 
\begin{equation}
\max \| \nabla_\bx \bu \| = \mathcal{0}(\delta), \ \delta \ll 1,
\label{eqn:max}
\end{equation}
we obtain 
\begin{equation}
\bE = \bfeps + \mathbf{0}(\delta^2), \  \ 
\be = \bfeps + \mathbf{0}(\delta^2), \  \
\bB = \bI + 2\bfeps + \mathbf{0}(\delta^2),
\label{eqn:max-2}
\end{equation}
where $\bfeps$ is the linearized strain defined as,
\begin{equation}
\bfeps := \frac{1}{2} \left( \nabla \bu +  \nabla \bu^{T}  \right),
\label{LinStrain}
\end{equation} 
where $( \; \cdot \; )^{T}$ is the \textit{transpose} operator.
Hence we can approximate $\bE (\bfa{x})$ by $\bfeps (\bfa{x})$ then there is no distinction between reference and current configuration and we can interchange $\bx$ and $\bX$. \\

Let $\bsigma$ denotes the Cauchy Stress tensor in the deformed configuration, then the first and second Piola-Kirchhoff Stress tensors in the reference configuration are given by
\begin{equation}
\bS:= \bsigma \bF^{-\text{T}} \text{det}(\bF) 
\ \text{ and }\ 
\bar{\bS}:= \bF^{-1} \bS,
\end{equation}
respectively. 
The material body is called Cauchy elastic if its constitutive class is determined by a response function of the relation 
$\bS = \hat{\bS}(\bF).$
Thus, the Cauchy stress $\bsigma$ is a function of the deformation gradient $\bF$, 
and the stress depends on the stress-free and final configurations of the body \cite{Truesdell2004}. 
For a compressible, homogeneous, isotropic elastic body, the Cauchy stress is given by:
\begin{equation}
\bsigma = \alpha_1 \bI + \alpha_2 \bB + \alpha_3 \bB^2,
\label{eqn:12}
\end{equation}
where $\alpha_i$, $i=1,2,3$ depend on isotropic invariants $\rho, tr (\bB), tr (\bB^2),$ and $tr (\bB^3)$, and $\rho$ is the density of the body \cite{Truesdell2004}. 
Next, the body is called Green elastic (or hyper elastic)  \cite{truesdell1955hypo}  
if the stress response function is the gradient of a scalar-valued potential, i.e 
$\hat{\bS}(\bF) = \partial_\bF \hat{w}(\bF),
$
and  a stored energy exists.
Thus, the stress in a Cauchy elastic body and the stored energy associated with a Green elastic body depend only on the deformation gradient
as discussed in \cite{carroll2009must}.

\subsubsection{Implicit and strain-limiting constitutive models}
\label{sec:nonlinear}

The general class of elastic materials that are far richer compared to classical Cauchy elastic bodies are introduced in a series of papers by Rajagopal and his co-authors in ~\cite{rajagopal2007elasticity,rajagopal2007response,rajagopal2009class,rajagopal2003implicit,rajagopal2011non,rajagopal2011conspectus,rajagopal2014nonlinear,bridges2015implicit,bustamante2014note}. 
In \cite{rajagopal2007elasticity},  it is assumed that Cauchy stress and stretch  are implicitly related by a relation of the type,
\begin{equation}
\mathcal{F}(\bsigma, \bB) = \mathbf{0}.
\label{eqn:19}
\end{equation}
A special subclass of Equation~\eqref{eqn:19} has 
the Cauchy stretch which is an explicit function of Cauchy stress and is given as
\begin{equation}
\bB = \tilde{\alpha}_1 \bI +  \tilde{\alpha}_2 \bsigma +  \tilde{\alpha}_3 \bsigma^2,
\label{eqn:23}
\end{equation}
where  $\tilde{\alpha}_i$, $i=1,2,3$ are the scalar-valued functions of the isotropic invariants of
\[
\rho, tr(\bsigma),tr(\bsigma^2),tr(\bsigma^3).
\] 
Under the linearization of Equation~\eqref{eqn:max}, the model (Equation~\eqref{eqn:23}) leads to 
\begin{equation}
\bfeps = \beta_1 \bI + \beta_2 \bsigma + \beta_3 \bsigma^2, 
\label{eqn:30}
\end{equation}
where one can see that the linearized strain $\bfeps$ is given as a nonlinear function of the stress $\bsigma$ and here $\beta_1$ is dimensionless and the material moduli $\beta_2$ and $\beta_3$ need to have dimensions that 
are the inverse of the stress and the square of the stress, respectively. 
{We note that the above relation for elastic bodies (Equation~\eqref{eqn:30}) has profound implications in studying stress-strain concentration near fracture tips in elastic materials. The relationship 
does not require stress to be ``small'' but  strains will be uniformly bounded throughout the body including {at the} tips of cracks and fracture. Hence, it predicts meaningful strain values near the crack-tips thereby removing unphysical strain singularity from LEFM model \cite{rajagopal2011modeling, kulvait2013, gou2015modeling,bridges2015implicit}. } In this paper, we only consider ``isotropic'' elastic bodies by means of definition given by Equation~\eqref{eqn:30} for simplicity.  A general linearization procedure to obtain models for anisotropic nonlinear elastic bodies defined by implicit relationship is given in \cite{gou2015modeling,Mallikarjunaiah2015,MalliPhD2015}.

Now, let us turn our attention to formulate a meaningful boundary value problem within the framework of strain-limiting nonlinear elastic models. To that end, we first consider an isotropic elastic material, in the absence of body coupling and body force, then the balance of linear and angular momentum  
reduces to 
\begin{equation}\label{eq:blaws}
-\nabla \cdot \bsigma =\bf{0}, \quad \mbox{and} \quad \bsigma = \bsigma^T.
\end{equation}
Further, the linearized strain tensor needs to satisfy the compatibility condition such as
\begin{equation}\label{eq:compcond}
\curl \, \curl \, \bfeps = \bf{0},
\end{equation}
where $\curl$ is classical ``curl'' operator for the second-order tensors. In this paper, we consider the problems within Equation~\eqref{eqn:30} and formulate the boundary value problem by introducing \textit{Airy's stress function}. We note that solving Equation~\eqref{eq:compcond} reduces into an elegant quasi-linear partial differential equation.
Thus, the system of partial differential equations that define the problem within the nonlinear elasticity is
\begin{subequations}
\begin{align}
-\nabla \cdot \bsigma &=\bf{0}, \quad \mbox{and} \quad \bsigma = \bsigma^T, \label{equilib:eq} \\
\bfeps &= \Psi_{0}\left( \tr \bsigma, \; \| \bsigma \|  \right) \bI + \Psi_{1}\left(  \| \bsigma \|  \right) \bsigma, \label{eqn:main} \\
\curl \, \curl \, \bfeps &=\bf{0}, \\
\bfeps &= \frac{1}{2} \left( \nabla \bu +  \nabla \bu^{T}  \right).
\end{align}
\end{subequations}
In Equation \eqref{eqn:main}, $\Psi_{0}(\cdot, \cdot), \; \Psi_{1}(\cdot)$ 
are scalar functions of stress invariants and more importantly the assumption of no residual stress implies $\Psi_{0}\left( 0, \; \cdot \right) = 0$.

\subsection{Anti-plane strain or Mode-III problem}
The problem considered in this work is the quasi-static crack evolution under anti-plane strain (or tearing) loading. The anti-plane shear is planar, meaning all kinematical quantities such as displacement vector $\bfa{u}(\bfa{x}, t) $,  stress tensor  $\bfsig(\bfa{x},t) $, and strain tensor  $\bfeps(\bfa{x},t) $ depend only upon the in-plane variables $x_{1}$ and $x_{2}$. In an anti-plane strain problem, the displacements in the body are zero, while the out-of-plane displacement is dependent on the in-plane co-ordinates $x_1$ and $x_2$ but independent of $x_3$. Therefore, the only non-zero component of the displacement vector is in $x_{3}$-direction, i.e.,
\begin{equation}\label{eq:disp_vector}
\bfa{u}(x_1,\, x_2,\, t) = \left( 0, 0, u( x_1,\, x_2,\, t)  \right).
\end{equation}
 Further, the only non-zero components of the stress tensor $\bfsig$ are $\sigma_{13}$ and $\sigma_{23}$. Then, the stress in the classical linearized isotropic elastic model depicted as
\begin{equation}\label{Eq:Constitutive}
\bfsig = 2 \, \mu \, \bfeps + \lambda \, \tr \left( \bfeps \right) \,  \bf{I},
\end{equation}
reduces to 
\begin{equation}
\bfsig = 2 \, \mu \, \bfeps,
\end{equation}
where $\mu$ and $\lambda$ are \textit{Lam$\acute{e}$} coefficients. It follows from Equation~\eqref{eq:disp_vector}  that the only non-zero components of the strain tensor $\bfeps$ are $\epsilon_{13}$ and   $\epsilon_{23}$.

Now, since $\tr (\bfsig) = 0$, the constitutive relationship of Equation~\eqref{eqn:main} takes the following form as
\begin{equation}
\bfeps = \Psi_{1}\left( \| \bsigma \|  \right) \bsigma. \label{eqn:main2}
\end{equation}
For the planar problem on hand with only two non-zero strain components, Equation~\eqref{eq:compcond} takes the form as
\begin{equation}\label{pde:strain}
\frac{\partial }{\partial x_2} \, \epsilon_{13} - \frac{\partial }{\partial x_1} \, \epsilon_{23} = 0.
\end{equation}
In order to derive partial differential equation, let us exploit the definition of \textit{Airy's stress function} $\Phi=\Phi(x_1, \; x_2)$ as 
\begin{equation}\label{eq:airy:strcomp}
\sigma_{13}:= \frac{\partial \Phi}{\partial x_2}, \quad  \sigma_{23}:= - \frac{\partial \Phi}{\partial x_1},
\end{equation}
which automatically satisfies the equilibrium equation (Equation~\eqref{equilib:eq}). Using Equation~\eqref{eq:airy:strcomp} in Equation~\eqref{eqn:main2}, we obtain
\begin{subequations}\label{str:nonlin}
\begin{align}
\epsilon_{13} &= \Psi_{1}\left( \| \nabla \Phi \|  \right) \Phi_{,2},  \\
\epsilon_{23} &= - \Psi_{1}\left( \| \nabla \Phi  \|  \right) \Phi_{,1}, 
\end{align}
\end{subequations}
and now using Equation~\eqref{str:nonlin} in Equation~\eqref{pde:strain}, we get a second-order quasi-linear partial differential equation
\begin{equation}\label{pde:nlin1}
  - \nabla \cdot \left( \Psi_{1} \left( \| \nabla \Phi \| \right) \;  \nabla \Phi \right) =0,
\end{equation}
with 
\begin{equation}\label{eqn:grad_airy_norm}
\|\nabla \Phi \|^2 = \left( \partial_{x_1}  \Phi \right)^2 + \left( \partial_{x_2}  \Phi \right)^2.
\end{equation}
In the reminder of this paper, we use the following particular form of the constitutive function $\Psi_{1}$, which a similar form has been used to study stress-strain near a static wedge \cite{kulvait2013,kulvait2019} and elliptical hole \cite{ortiz2012},  
\begin{equation}\label{eq:Psi}
 \Psi_{1} (\|\bfsig\|) = \frac{1}{2 \, \mu \left( 1 + \beta^\alpha \, \| \bfsig \|^{\alpha} \right)^{1/\alpha}},
\end{equation}
where the positive constants $\beta$ and $\alpha$ are modeling parameters. In the view of Equation~\eqref{eq:Psi}, the nonlinear PDE (Equation~\eqref{pde:nlin1}) now takes the form as
\begin{equation} \label{pde:mech}
- \nabla \cdot \left( \frac{\nabla \Phi}{2 \, \mu \left( 1 + \beta^\alpha \; \|\nabla \Phi \|^{\alpha} \; \right)^{1/\alpha}}  \right) = 0.
\end{equation}
{We emphasize that the above nonlinear equation allows a remarkable departure from the classical singularity of strains near the crack-tip even though stress is allowed to be singular. Thus, we aim to augment the model with the local critical crack-tip fracture criterion as in \cite{BourFraMar00,BourFraMar08,HeWheWi15,LeeWheWi16} to study the quasi-static crack evolution. 
}
\begin{remark}
It is very clear that the nonlinear elastic material model presented in Equation~\eqref{eqn:30} is hyperelastic and has the corresponding complementary 
and strain energy functions associated with it. For a special case with $\alpha = 1$, the constitutive relation of Equation~\eqref{eqn:main2} takes the form
\begin{equation}
\bfeps = \frac{\bfsig}{1 + \beta \, \| \bfsig \| },
\end{equation}
and the inverted constitutive relationship for ``stress'' as a nonlinear function of linearized ``strain'' is given by 
\begin{equation}
\bfsig = \frac{\bfeps}{1 - \beta \, \| \bfeps \| }.
\end{equation}
One can also derive the ``stress'' by a scalar strain energy function, i.e.,
\begin{equation}\label{eq:stress-energy}
\bfsig = \frac{\partial \Xi(\|  \bfeps  \| )}{\partial \bfeps},
\end{equation}
where $\Xi (\|  \bfeps  \| )$ is the associated strain energy function and for $\alpha = 1$, it is given by
\begin{equation}
\Xi ( \|  \bfeps  \| ) := \frac{1}{\beta} \left(  \log(1- \beta \, \|  \bfeps  \| )  + \beta \, \|  \bfeps  \|  \right).
\end{equation}
To derive the associated strain energy function from Equation~\eqref{eq:stress-energy} when $\alpha \neq 1$, one can use hyper-geometric functions. 
\end{remark}

\subsection{Quasi-static evolution and phase-field regularization}
Let $\Lambda:=\Lambda(t) \in \mathbb{R}^d\;(d=2, 3)$ be a smooth, open, connected, bounded domain with a given boundary $\partial \Lambda$. It contains a set $\Gamma(t) \in \mathbb{R}^{d-1}$, a lower dimensional crack set across which the displacements suffer discontinuity. Here, the time is denoted by $t\in[0,T]$, with the final time $T>0$ in the computational time interval. 
We assume that the discontinuity set $\Gamma(t)$ is completely contained within $\Lambda(t)$, and $\Gamma(t)$ is a Hausdorff measurable set. 
The energy functional established in \cite{FraMar98} describes the total energy of the material body given by 
\begin{equation}\label{eq:tenergy}
{E}(\Phi, \; \Gamma):=\int_{\Lambda \setminus \Gamma} \mathcal{W}(\Phi) d\bfx + G_c \mathcal{H}^{d-1}(\Gamma),
\end{equation}
where $\mathcal{W}(\cdot) \colon H^{1}(\Lambda) \to \mathbb{R}$ is the elastic energy, $\Phi \colon \Lambda \to \mathbb{R}$ is Airy's stress function, $\mathcal{H}^{d-1}$ denotes Hausdorff measure, and $G_{c}>0$ denotes the critical energy release rate of material (or fracture toughness).
 The total energy, defined via Equation~\eqref{eq:tenergy}, is the balance between stored elastic energy of the material and the crack-surface energy {needed to create new increments of crack}. 
Then the unilateral minimization of the total energy yields a new equilibrium (in the sense of Griffith) and a new crack set, which might result propagation of the given crack.  The minimization process is labelled as unilateral because the unknown crack set $\Gamma$ cannot decrease in time.  The time-dependent minimization of the above energy functional has been studied extensively for the existence of solutions by using the method of calculus of variations, for linear elasticity  \cite{dal2002model, francfort2003existence}  and for finite elasticity \cite{dal2005quasistatic}.

\begin{figure}[!h]
\centering
\includegraphics[width=0.7\textwidth]{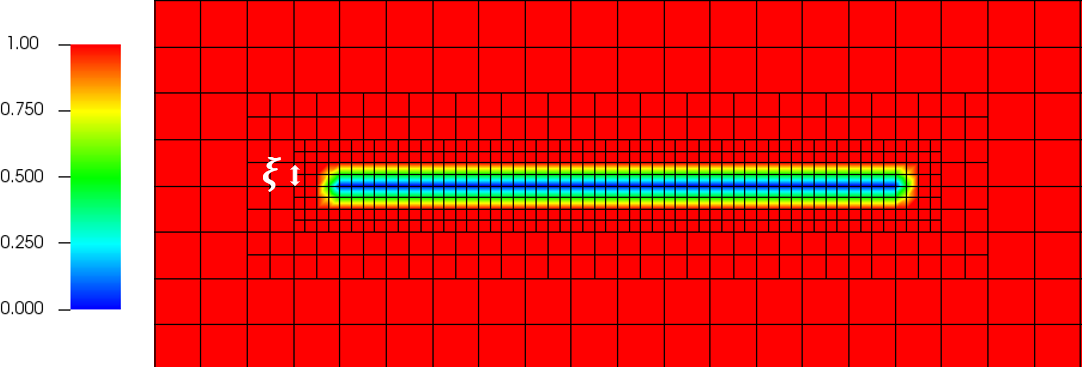}
\caption{An example of a fracture defined with the phase-field function $\varphi \in [0,1]$ expressed with a regularization parameter $\xi$  and adaptive meshes.}
\label{fig:figure_1}
\end{figure}

To consider a regularization of the total energy that can be readily implementable by standard finite element techniques,
Ambrosio-Tortorelli energy functional  \cite{AmTo90, AmTo92}, $E_{\xi} \colon \mbox{H}^{1} \left( \Omega; \mathbb{R} \right) \times \mbox{H}^{1} \left( \Omega; [0, \; 1] \right) \to \mathbb{R}$, 
is introduced as following

\begin{equation}\label{reg:energy}
E_{\xi} (\Phi, \; \varphi) := \frac{1}{2} \int_{\Omega} \left( (1- \kappa) \varphi^2 + \kappa  \right) \; \mathcal{W}(\Phi) \; d \bfx + G_c \int_{\Omega} \left[ \frac{(1-\varphi)^2}{2\xi} + \frac{\xi}{2} \; |\nabla \varphi|^2  \right]  \; d\bfx,
\end{equation}
where $\varphi \in \mbox{H}^{1} \left( \Omega; [0, \; 1] \right) $ is a scalar \textit{phase-field function}, 
$\kappa \ll 1$ is a numerical regularization parameter~\cite{MiWheWi15b} for the bulk energy term.
Here, the $\varphi=0$ indicates the fracture zone and $\varphi=1$ defines the non-fractured zone, where
$\xi > 0$ is a regularization parameter which is the critical length of the diffusive zone ($\varphi \in (0,1)$) for phase-field variable $\varphi$. See Figure \ref{fig:figure_1} for more details. 
Thus, the energy considers only the crack energy if $\varphi=0$ since the bulk energy vanishes (by assuming $\kappa\approx 0$). On the other hand, only the bulk energy is considered if $\varphi=1$ since the fracture energy is zero. 
We have both nonzero bulk and fracture energies interpolated in the diffusive zone. 
Moreover, the above energy functional Equation~\eqref{reg:energy} will be minimized with the irreversibility condition, $\partial_{t} \varphi \leq 0$.  The latter condition is where we only allow the crack to propagate (but not bonding), and the crack evolution  is formulated in terms of quasi-static assumptions.

Finally, the system is supplemented by the time-dependent non-homogeneous Dirichlet boundary conditions applied on the part of the boundary $\partial \Lambda_{D}$ and Neumann boundary conditions on the rest of the boundary $\partial \Lambda_{N}$, 
thus  $\partial \Lambda_{D} \cup \partial \Lambda_{N} = \partial \Lambda$, and $\partial \Lambda_{D} \cap \partial \Lambda_{N} =\emptyset$. 
The boundary conditions of the elastic energy depend on the setup of  problem \cite{SogoLeeWheeler_2018}, {and} we employ homogeneous Neumann boundary condition for the phase-field.

\begin{remark}
The sequence of functionals $\left\{  E_{\xi} \right\}_{\xi > 0}$ defined in Equation~\eqref{reg:energy}, for linear elasticity, 
is known to have $\Gamma$-convergence \cite{braides2002gamma} to $E$  as in Equation~\eqref{eq:tenergy} in $\mbox{L}^{1}(\Omega) \times \mbox{L}^{1}(\Omega)$ as $\xi \to 0$. The existence of minimizers for $E_{\xi}$ has been shown in \cite{AmTo92} for each $\xi, \; \kappa >0$. The role of one regularization parameter $\kappa$ is to regularize the bulk (or strain) energy. This parameter needs to be small, and should not change, in the entire computation in order to avoid a over-estimation of the bulk energy which results in an under-estimation of the crack-surface energy. In \cite{BourFraMar00}, $\kappa$ was tied to the value of $\xi$ so that $\Gamma$-convergence results are valid, where as $\kappa$ was kept zero in the original $\Gamma$-convergence result. For the quasi-static problem, it is still an open issue about the choice of the parameter $\kappa$ that yields physically meaningful crack pattern and corresponds to a particular experiment. 
\end{remark}

\section{Numerical Method}
\label{sec:num}

In this section, we present a finite element method utilized for the spatial discretization of the coupled nonlinear mechanics and phase-field system. In addition, the decoupling algorithm between the elasticity and the phase-field equations, so called the L-scheme  \cite{brun2019iterative} is presented. Finally, the Euler-Lagrange formulation for our governing system with the augmented Lagrangian method \cite{fortin2000augmented,glowinski1989augmented,wheeler2014augmented}  for the irreversibility constraint, and the linearization of the given nonlinear problems are discussed. 
We start with the temporal discretization which considers the quasi-static fracture propagation with the irreversibility condition.

\subsection{Temporal Discretization}
First, we define a partition of the time interval $0=:t^0 <t^1 < \cdots <  t^N := \mathbb{T} $ and denote the uniform time step size by $\Delta t:= t^n - t^{n-1}$. 
Then, we denote the temporal discretized solutions by 
\begin{equation}
\Phi^n := \Phi(\cdot, \, t^n) \ \ \text{ and } \ \ 
\varphi^n := \varphi(\cdot, \, t^n).
\end{equation}
The crack-irreversibility condition for the phase-field variable, 
\begin{equation}
\partial_{t} \varphi \leq 0,
\end{equation}
is discretized by 
\begin{equation}
\varphi^n \leq \varphi^{n-1}, 
\end{equation}
by employing a backward Euler discretization. 
In this paper, we employ a simple straight-forward penalization technique to accommodate the crack-irreversibility condition.
To that end, we define and add the following penalty term
\begin{equation}
P(\gamma,\varphi^n) := [\lambda + \gamma \, \bar{\varphi}^n \, ]^+
\end{equation}
where $\bar{\varphi}^n  = (\varphi^n - \varphi^{n-1})$ and $\gamma$ is the penalization parameter. The subscript $[\cdot]^+$ denotes the positive part of a function, i.e 
\[
[f]^+ = \max(0,f).
\]
For a better performance, we utilize the augmented Lagrangian method \cite{fortin2000augmented,glowinski1989augmented,wheeler2014augmented} by  adding a function $\lambda \in L^2(\Lambda)$ which is given and updated through the iterations. 
Then, we rewrite the definition of the  total energy  which  includes the penalization term as
\begin{align}\label{eq:energy_min}
E_{\xi}(\Phi^n, \varphi^n) = & \frac{1}{2} \int_{\Lambda} ((1-\kappa)(\varphi^n)^2 + \kappa) \,  \mathcal{W}(\Phi^n) \, d\bfx 
+ G_c \int_{\Lambda} \left( \frac{(1-\varphi^n)^2}{2 \xi} + \frac{\xi}{2} \, |\nabla \varphi|^2 \right) \, d\bfx \notag \\
&+ \dfrac{1}{2 \gamma} \| P(\gamma, \, {\varphi}^n) \|^2. 
\end{align}
 The term $\mathcal{W}(\Phi^n)$ in Equation~\eqref{eq:energy_min} is the \textit{bulk} or \textit{strain  energy}, and can be obtained by 
 \begin{equation}
 \mathcal{W}(\Phi) := \bfsig \colon \bfeps =  \frac{ \| \nabla \Phi^n \|^2}{2 \, \mu \left( 1 + \beta^\alpha \; \| \nabla \Phi^n \|^{\alpha} \; \right)^{1/\alpha}}, \label{senergy}
 \end{equation}  
as discussed in the previous section. 
Thus, we solve the above constrained energy minimization problem to seek the scalar-valued \textit{Airy's stress function} $\Phi$  and the scalar-valued phase-field variable $\varphi$.

\subsection{Spatial Discretization}
We consider a mesh family $\{\mathcal{T}_h \}_{h>0}$, which is assumed to be shape regular in the sense of Ciarlet, and we 
assume that each mesh $\mathcal{T}_h$ is a subdivision of $\bar{\Lambda}$ made of disjoint elements $\mathcal{K}$, i.e., squares when  $d=2$ or cubes when $d=3$. 
Each subdivision is assumed to exactly approximate the computational domain, thus $\bar{\Lambda} = \cup_{K\in\mathcal{T}_h} \mathcal{K}$. 
The diameter of an element $\mathcal{K}\in \mathcal{T}_h$ is denoted by $h$
and we denote $h_{\min}$ for the minimum. For any integer $k \geq 1$ and any
$\mathcal{K} \in \mathcal{T}_h$, we denote by $\mathbb{Q}^k(\mathcal{K})$ the
space of scalar-valued multivariate polynomials over $\mathcal{K}$ of partial
degree of at most $k$.

In this section, we present a fully-coupled Euler-Lagrange formulation 
for $\Phi_h$ and $\varphi_h$, approximating Airy's stress and phase-field, $\Phi, \varphi$, respectively. We consider a time-discretized system in which 
time enters through the irreversibility condition. 
Let $V_h \times W_h \subset V \times W$ be the discrete space formulated by continuous Galerkin approximations. 
The spatial  discretized solution variables are 
$\Phi_h \in \mathcal{C}^1([0,T];V_h(\mathcal{T}))$ and 
$\varphi_h \in \mathcal{C}^1([0,T];W_h(\mathcal{T}))$, 
where
\begin{align}
&V_h(\mathcal{T}) := 
\{ Y \in C^0(\bar{\Lambda};\mathbb{R}^d) \ | \ Y= 0 \ \text{on } \partial \Lambda, Y|_{\mathcal{K}} \in {\mathbb{Q}}^1(\mathcal{K}), \forall \mathcal{K} \in \mathcal{T} \} , \\
&W_h(\mathcal{T}) := 
\{ Z \in C^0(\bar{\Lambda};\mathbb{R}) | 
\ Z^{n+1}\leq Z^n \leq 1, Z|_{\mathcal{K}} \in \mathbb{Q}^1(\mathcal{K}), \forall \mathcal{K} \in \mathcal{T} \}.
\end{align}
For our convenience, from here on, we omit the $h$-subscript for $\Phi_h$ and $\varphi_h$ since we only consider the discrete solutions from now.

Next, we formulate the Euler-Lagrange equations and the finite element discretizations for the variational form of the energy functional 
$E_{\xi}(\Phi^n,\varphi^n)$ in Equation~\eqref{eq:energy_min}. We seek $U^{n}:=\{\Phi^n, \varphi^n \} \in V_h \times W_h$ such that 
\begin{multline}
\mathcal{L}(U^n)(\psi) =
(((1-{\kappa}) (\varphi^n)^2 + {\kappa})~ \mathcal{W}(\Phi^n), \nabla w)
 - 
G_c (  \dfrac{1}{\xi}  (1-\varphi^n) , \psi) \\
 +
G_c ( {\xi} \nabla \varphi^n ,\nabla \psi) =0,  
\ \ \forall w, \psi \in \Psi := \{ w, \psi \} \in V_h \times W_h,
\label{eqn:1}
\end{multline}
for each $t^n$. 
For the simplicity, we define the degradation function with the phase-field function as 
\begin{equation}
g({\varphi}) := ((1-{\kappa}) (\varphi^n)^2 + {\kappa}).
\label{eqn:g_phi}
\end{equation}
Then, by computing a directional derivative of Equation~\eqref{eqn:1} with respect to $\Phi$ and $\varphi$, we obtain the following subproblems 
\begin{equation}
\mathcal{L}_1(\Phi^n, w)
:= (g({\varphi}) ~\mathcal{W}(\Phi^n) , \nabla w) = 0,
 \ \ \forall w \in V_h,
\label{sform1}
\end{equation}
and 
\begin{multline}
\mathcal{L}_2(  \varphi^n, \psi)
:=
(1-{\kappa})(   \varphi^n \mathcal{W}(\Phi^n) , \psi )
 -
 G_c (  \dfrac{1}{\xi} ( 1-   \varphi^n )  , \psi) \\
 +
 G_c ( {\xi} \nabla  \varphi^n ,\nabla \psi) 
 + ( [\lambda + \gamma(\varphi^n - \varphi^{n-1})]^+, \psi) =0, 
\ \ \forall  \psi \in W_h.
\label{sform2}
\end{multline}   
Here, we denote $\mathcal{L}_1$ as the mechanics subproblem and  $\mathcal{L}_2$ as the phase-field subproblem.

\subsection{Iterative Algorithm}

For each time step $n$, the iterative algorithm defines a sequence 
$\{ \Phi^{n,i}, \varphi^{n,i} \}$, where $i \geq 0$ denotes each iteration steps. The iteration is formulated with two steps. 
First the mechanics subproblem $\mathcal{L}_1$ (Equation~\eqref{sform1}) is solved with the given phase-field and its degradation function and {Airy's stress value} from the previous iteration, i.e $\{ \Phi^{n,i-1}, \varphi^{n,i-1} \}$. 
In the first iteration ($i=1$), we set $\Phi^{n,i-1}=\Phi^{n,0}:=\Phi^{n-1}$ ($\varphi^{n,i-1}=\varphi^{n,0}:=\varphi^{n-1}$).
Then, the phase-field subproblem $\mathcal{L}_2$ (Equation~\eqref{sform2}) is solved with the known bulk energy function computed using {Airy's stress function} from the previous iteration. 
This iterative algorithm, the staggered L-scheme, which was introduced in \cite{brun2019iterative}.
We note that there are two positive stabilization constant terms $L_{\Phi}$ and $L_{\varphi} $ which depend on the problem. 
Moreover, each nonlinear subproblem, {$\mathcal{L}_1$ and $\mathcal{L}_2$}, is linearized by utilizing the Newton's method. 
For the faster convergence of our nonlinear problem, we note that the linear (elasticity) problem is employed as a initial guess at the initial iteration and previous solution is used in subsequent iterations as old solution.

\subsubsection{Step 1. Solve the mechanics subproblem}
For the L-scheme iteration between mechanics and phase-field subproblems,  $i=0,\, 1, \, 2,\ldots$,  
we first solve for $\Phi^n \in V_h$ with given $\Phi^{n-1}, \varphi^{n-1}$ satisfying
\begin{equation}
\mathcal{L}_1(\Phi^{n, i}, \, w) = 0, \ \ \forall w \in V_h,
\label{eqn:MNL}
\end{equation}
where
\begin{equation}
\mathcal{L}_1( \Phi^{n, i}, \, w)
:= (g(\varphi^{n, \,i-1})  \mathcal{W}( \Phi^{n,i} ) , w)
+ 
L_{\Phi}(\Phi^{n,i}   - \Phi^{n,i-1}, w).
\end{equation}
Here, the last term is an additional term from the L-scheme iterative method \cite{brun2019iterative} with a given positive parameter $L_\Phi$.

To solve the nonlinear problem, Equation~\eqref{eqn:MNL}, we employ the Newton iteration. 
Thus, we seek $\delta \Phi^{n,i} \in V_h$ by solving 
\begin{equation}
\mathcal{L}^{\prime}_1(\Phi^{n,i,a-1}, { \varphi^{n,i-1}})( \delta \Phi^{n,i,a}, w) = -\mathcal{L}_1(\Phi^{n,i,a-1})(w), \ \ \forall w \in V_h,
\end{equation}
for the Newton iteration steps $a=0,1,2,\ldots$ until 
$\| \delta \Phi^{n,i,a}\| \leq {\varepsilon_{\Phi}}$.
Then the Newton update is given by 
\begin{equation}
\Phi^{n,i,a} = \Phi^{n,i,a-1} + \omega_{\Phi} \delta \Phi^{n,i,a} ,
\end{equation}
where $\omega_\Phi$ is a line search parameter $\omega_\Phi \in [0,1]$. 
If the Newton iteration converges, we set 
\begin{equation}
\Phi^{n,i} = \Phi^{n,i,a}.
\end{equation}
Here, the Jacobian of $\mathcal{L}_1(\Phi(w))$ is computed as
\begin{equation}
\mathcal{L}^{\prime}_1(\Phi^{n,i}, \varphi^{n,i-1}) (\delta \Phi^{n,i,a},  w)\\ 
 := (  g(\varphi^{n,i-1}) \hat{\mathcal{W}}(\delta \Phi^{n,i,a} ) , \nabla w)
+ 
L_{\Phi}( \delta \Phi^{n,i,a}, w)
\end{equation}
where
\begin{multline}
\hat{\mathcal{W}}(\delta \Phi^{n,i,a} ) 
 :=
  \frac{\nabla \delta \Phi^{n,i,a} }{\left( 1 + \beta^\alpha \;  \| \nabla \Phi^{n,i,a-1} \|^{\alpha} \; \right)^{1/\alpha} } \\ -\frac{ \beta^{\alpha} \, (\nabla \Phi^{n,i,a-1} \cdot \nabla \delta  \Phi^{n,i,a}) \|\nabla \Phi^{n,i,a-1} \|^{\alpha -2} \, \nabla \Phi^{n,i,a-1} }{\left( 1 + \beta^\alpha \;  \| \nabla \Phi^{n,i,a-1} \| ^{\alpha} \; \right)^{1/\alpha + 1}}, 
\end{multline}
and 
\begin{multline}
\mathcal{L}_1( \Phi^{n,i}, w)
:= (((1 - \kappa) ( \varphi^{n,i-1})^2 + \kappa) \;  \bar{\mathcal{W}}( \Phi^{n,i,a-1} ) , \nabla w)
+ 
L_{\Phi}( \Phi^{n,i,a-1} -\Phi^{n,i-1}, w),
\end{multline}
where 
\begin{equation}
\bar{\mathcal{W}}( \Phi^{n,i,a-1} ) = 
\Bigg(   \frac{  \nabla \Phi^{n,i,a-1}}{ \left( 1 + \beta^\alpha \; \| \nabla \Phi^{n,i,a-1}   \|^{\alpha} \; \right)^{1/\alpha}}  \Bigg).
\end{equation}

\subsubsection{Step 2. Solve the phase-field subproblem}

Secondly,  we seek for $\varphi^{n,i} \in W_h$ satisfying 
\begin{equation}
\mathcal{L}_2(\varphi^{n,i}, \, \psi; \, \Phi) = 0, \ \ \forall \psi \in W_h,
\label{eqn:main_20}
\end{equation}
where
\begin{multline}
\mathcal{L}_2(\varphi^{n,i}, \, \psi; \, \Phi)
:=
(1-\kappa)(  \varphi^{n,i} \; \mathcal{W}(\Phi^{n,i}), \, \psi )
 -
 G_c (  \dfrac{1}{\xi} \, ( 1 -  \varphi^{n,i})  , \, \psi) 
 +
 G_c ( \xi \, \nabla  \varphi^{n,i} ,  \, \nabla \psi)  \\
 + (\eta^{n,i} (\lambda^{n,i} + \gamma ( \varphi^{n,i} - \varphi^{n-1} )), \, \psi) 
 +
 L_\varphi( \varphi^{n,i} - \varphi^{n,i-1}  ,\psi).
\label{eqn:main20}
\end{multline}
Here the last term is the L-scheme stabilization term with a positive constant value $L_\varphi$, and $\eta^{n,i} \in L^\infty(\Lambda)$ is defined as 
$$
\eta^{n,i}(x) :=
\begin{cases}
1, \ \ \text{ if } \lambda^{n,i}(x) + \gamma (\varphi^{n,i}(x) - \varphi^{n-1}(x)) > 0 \\
0, \ \ \text{ if } \lambda^{n,i}(x) + \gamma (\varphi^{n,i}(x) - \varphi^{n-1}(x)) \leq 0 \\
\end{cases}
$$
to replace the operator $[\cdot]^+$ with the computable form. 

To solve the nonlinear problem, Equation~\eqref{eqn:main_20}, we employ the Newton's method. To this end, we find $\delta \varphi^{n,i} \in W_h$ by solving 
\begin{equation}
\mathcal{L}^{\prime}_2(\varphi^{n,i,b-1})(\delta \varphi^{n,i,b}, \psi) = -\mathcal{L}_2(\varphi^{n,i,b-1})(\psi), \ \ \forall \psi \in V_h,
\end{equation}
for the Newton iteration steps $b=0,1,2,\ldots$, 
until $\| \delta \varphi^{n,i,b}\| \leq {\varepsilon_{\varphi}}$. 
Then we update 
$$
\varphi^{n,i,b} = \varphi^{n,i,b-1} + \omega_\varphi \delta \varphi^{n,i,b}.
$$
Here the Jacobian of $\mathcal{L}_2(\varphi(\psi))$ applied to a direction $\delta \varphi$ is 
\begin{multline}
\mathcal{L}^{\prime}_2(\varphi^{n,i,b-1})( \delta \varphi^{n,i,b}, \, \psi) := 
(1- \kappa)( \delta \varphi^{n,i,b}  \mathcal{W}(\Phi^{n,i}), \, \psi )
 +
 G_c ( \dfrac{1}{\xi} \delta \varphi^{n,i,b}  , \, \psi) \\
 +
 G_c ( {\xi} \nabla \delta \varphi^{n,i,b} , \, \nabla \psi) 
 + \eta^{n,i} \gamma( \delta \varphi^{n,i,b} , \, \psi) 
 +
 L_\varphi( \delta \varphi^{n,i,b} , \, \psi),
\end{multline}
and 
\begin{multline}
\mathcal{L}_2( \varphi^{n,i,b-1}, \, \psi)
:=
(1- \kappa)( \varphi^{n,i,b-1}  \mathcal{W}(\Phi^{n,i}), \, \psi )
 -
 G_c (  \dfrac{1}{\xi} ( 1-  \varphi^{n,i,b-1})  , \, \psi) 
 +
 G_c ( {\xi} \nabla  \varphi^{n,i,b-1} , \, \nabla \psi) \\
 + (\eta^{n,i} (\lambda^{n,i} + \gamma ( \varphi^{n,i,b-1} - \varphi^{n-1} )), \psi) 
 +
 L_\varphi( \varphi^{n,i,b-1} - \varphi^{n,i-1}  ,\psi).
\end{multline}   
If the Newton iteration converges, then we set 
$$\varphi^{n,i} = \varphi^{n,i,b}.$$
We also note that the augmented penalty term, 
$[ \lambda^{n,i-1} + \gamma (  \varphi^{n,i,b-1} - \varphi^{n-1} ) ]^+$
is updated every staggered steps.

Finally,  we employ both mechanics subproblem residual 
$\|\mathcal{L}_1(\Phi^{n,i},w)\|\leq \text{T}_{\text{OL}}$
and
phase-field subproblem residual  
$\|\mathcal{L}_2(\varphi^{n,i},\psi)\| \leq \text{T}_{\text{OL}}$ 
as the stopping criteria for both the L-scheme and augmented Lagrangian. 
If the whole iteration converges, we obtain
$$
 \Phi^{n} = \Phi^{n,i,a}  \ \ \  \text{ and } \ \ \ 
\varphi^{n} = \varphi^{n,i,b}.
$$

\section{Numerical Examples }
\label{sec:examples}

In this final section, we present several  numerical examples to verify and validate the  proposed nonlinear algorithm.
Moreover, we illustrate the capabilities and the effectiveness of the framework. All the computations for the nonlinear strain-limiting (or denoted as NLSL) model are developed by the authors based on the previous studies \cite{brun2019iterative,wheeler2014augmented}.  
The code is based on an open-source finite element package deal.II \cite{dealII90}, and all experiments utilize High Performance Computing (HPC)
at Texas A\&M University - Corpus Christi.

In Table \ref{Tab:Exs-1234}, the common parameters for the algorithm and numerical experiments for this section are presented.
In addition,  the phase-field regularization parameters are set as $\kappa=10^{-10}$ $h_{\min}$ and $\xi=$2$h_{\min}$, 
where $h_{\min}$ is the minimum cell diameter.
\begin{table}[!h]
\small
\begin{center}
\caption{Common parameters for all examples: Example 1 to Example 4.}
\label{Tab:Exs-1234}
\begin{tabular}{cl}
\toprule
\textbf{Parameter} & \textbf{Value}\\
\midrule
Tolerance for the Newton Iteration ($\varepsilon_{\Phi}$, $\varepsilon_{\varphi}$) & 1.0e-7\\
L-scheme  coefficients $L_\Phi$, $L_\varphi$ & 1.0e-6 \\
Tolerance for the L-scheme and the augmented Lagrangian Iteration (T$_{\text{OL}}$) & 1.0e-6 \\
\bottomrule
\end{tabular}
\end{center}
\end{table}

\subsection{Example 1: Convergence tests}

We  first verify our implementation of the proposed algorithm in previous section  for the nonlinear Airy's stress equation~$\mathcal{L}_1$ 
(Equation~\eqref{eqn:MNL}) by presenting the optimal error convergence. 
For the simplicity, only the nonlinear mechanics subproblem is considered and the phase-field variable is neglected for this example. Thus, we set the phase-field value to be ``1'' for the whole domain  and ${\kappa=0}$. 

\begin{table}[!h]
\centering
\small
\begin{tabular}{|c|l||l|l||l|l|}
\hline
\multirow{2}{*}{DOF} & \multicolumn{1}{c||}{\multirow{2}{*}{$h$}} & \multicolumn{2}{c||}{LEFM}                               & \multicolumn{2}{c|}{NLSL}                            \\ \cline{3-6} 
                       & \multicolumn{1}{c||}{}                   & \multicolumn{1}{c|}{L2 Error} & \multicolumn{1}{c||}{Rate} & \multicolumn{1}{c|}{L2 Error} & \multicolumn{1}{c|}{Rate} \\ \hline\hline
9                      & 0.25                                    & 0.250000000000                & 0.0                       & 0.206592351198                & 0.0                       \\ \hline
25                      & 0.125                                   &  0.067876629531                 & 2.6942                    & 0.059590231627                & 2.4338                  \\ \hline
81                      & 0.0625                                  & 0.017249573022               & 2.3542                    &0.015062531456                & 2.3398                    \\ \hline
289                      & 0.03125                                 &  0.004329234362               & 2.1788                    & 0.003735017497              &2.1926                  \\ \hline
1089                      & 0.015625                                & 0.001083351206               & 2.0898                    &  0.000921668019           & 2.1097                   \\ \hline
4225                     & 0.0078125                               & 0.000270902819              & 2.0450                    &  0.000226948716               & 2.0674                     \\ \hline
\end{tabular}
\caption{Example 1. The results of $L^2$ error convergence test of the approximated Airy's stress variable for the linear {(LEFM)} and the nonlinear {(NLSL)}  mechanics subproblems  are illustrated. 
For the linear case, 
the parameters are set to $\beta = 0$, but $\alpha = 1$ and $\beta = 0.2$ for the nonlinear case. We observe the optimal convergence for both cases.}
\label{fig:ex1}
\end{table}

The exact solution for the mechanics subproblem is chosen as 
\begin{equation}\label{eq:Ex0}
\Phi(x,y) := \sin{(\pi x)}\sin{(\pi y)},
\end{equation}
in the computational domain $\Lambda = [0,1]^2$.
The right hand side and the boundary conditions are chosen accordingly to satisfy the homogeneous boundary conditions on $\partial \Lambda$. 
In addition, the shear modulus is set as {$\mu=0.01$} and 
the nonlinear parameters are given as {\color{black}$(\alpha, \beta)=(1,0)$} for the linear case and  {\color{black}$(\alpha, \beta)=(1,0.2)$} for the nonlinear case. 
Six computations on the uniform meshes were computed where the mesh size $h$ is divided by two for each cycle. 
The corresponding number of degrees of freedom (DOFs)  for each cycle is $9, 25, 81, 289, 1089$, and $4225$. The results of $L^2(\Lambda)$ errors for the approximated solution 
for each mesh size $h$ are shown in 
Table~\ref{fig:ex1}. The $L^{2}$ error depicted in the table, for both LEFM and NLSL problems, is optimal since we use $Q_1$ finite elements.

\subsection{Example 2: Static crack and the strain-limiting effect}
\begin{figure}[!h]
\centering
\centering
\begin{tikzpicture}
\draw (0,0) -- (3,0) -- (3,3) -- (0,3) -- (0,0);
\draw [line width=0.5mm, blue]  (1.5,1.5) -- (3,1.5);
\draw [loosely dotted, line width = 0.5mm, red] (0,1.5) -- (1.5,1.5);
\node at (-0.3,-0.25)   {$(0,0)$};
\node at (3.3,3.2)   {$(1,1)$};
\node at (-0.3, 1.5)   {$\Lambda_L$};
\node at (3.3, 1.5)   {$\Lambda_R$};
\node at (1.5, -0.25)   {$\Lambda_{B}$};
\node at (1.5, 3.2)   {$\Lambda_{T}$};
\node at (2.2, 1.8)   {$\Gamma_{C}$};
\node at (3.5, 2.0) {$\bigodot$};
\node at (3.5, 2.7) {$\bigodot$};
\node at (3.5, 0.3) {$\bigotimes$};
\node at (3.5, 1.0) {$\bigotimes$};

\node at (4.1, 2.35) {$\Phi =  c$};
\node at (4.2, 0.65) {$\Phi =  -c$};
\end{tikzpicture}
\includegraphics[width=0.4\textwidth]{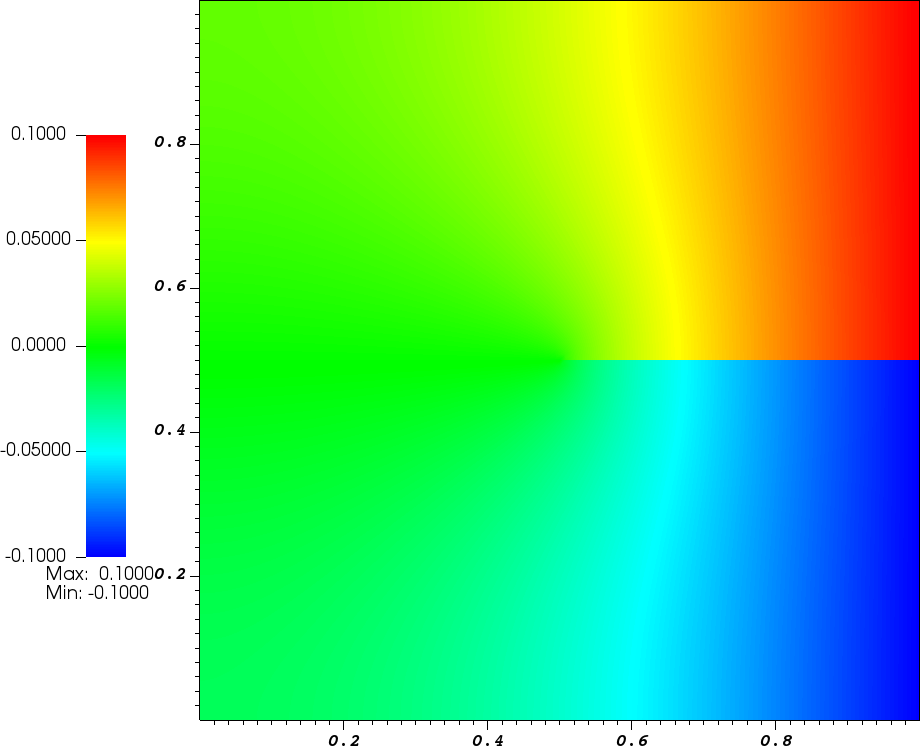}
\caption{Example 2 and 3. (Left) A setup with the boundary conditions. 
The bold blue line (0.5,0.5)-(1, 0.5) denotes the slit and red-dotted line is the expected crack path. 
The traction for the anti-plane shear is applied on the $\Lambda_{R}$. (Right) Computed Airy's stress values for LEFM case. }
\label{Fig:ex2_setup}
\end{figure}
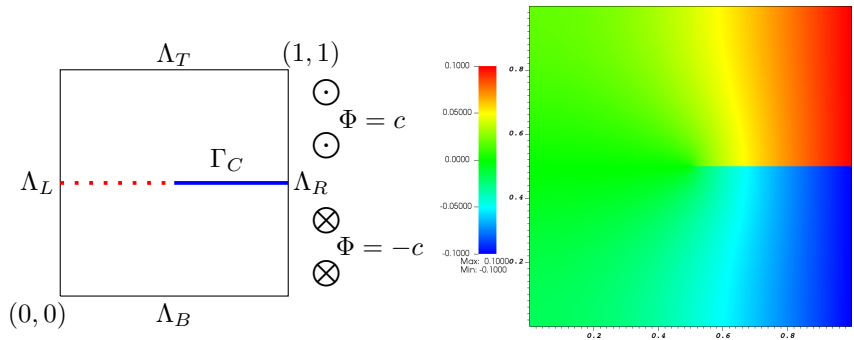

From Example 2 to all the succeeding examples, the anti-plane shear problems are introduced, and Example 2 and 3 share the same computational domain and the boundary conditions as illustrated (Left) in Figure~\ref{Fig:ex2_setup}.  
We consider the traction applied on the top and bottom half of the right boundary ($\Lambda_R$). 
The tractions are imposed through the Dirichlet boundaries {as $\Phi = c$ and $\Phi = -c$ for the  top and bottom half of the right boundary, respectively. Here, $c$ is a positive constant.}  
The remaining boundary parts including the slit boundary ($\Gamma_C$) are kept traction free. More detailed information regarding the suitable boundary conditions for $\Phi$ are discussed in  \cite{kulvait2013,kulvait2019}.
In addition, the physical parameters are given as
shear modulus $\mu=1.0$ $Pa$ and critical energy release rate $G_c= 0.01$ $Nm^{-1}$.

The domain is refined globally seven times resulting in the minimum cell diameter as $h_{min}= 0.0110485$. 
The Dirichlet boundaries are set with a condition $c=0.1$.
The nonlinear model parameter $\alpha$ is set to $\alpha=1.5$ and we test several different $\beta$ values {including $\beta=1.0, 2.0, 5.0, 10.0$ and $25.0$} to present the strain-limiting effects. 
We focus and compare the stress and strain values over the center line $(0, 0.5)-(0.5,0.5)$ which lead up to the crack-tip as illustrated as a red-dotted line in Figure~\ref{Fig:ex2_setup}.

{Figure \ref {Fig:ex2_setup} (Right) presents the computed Airy's stress values over the domain. The discontinuity across the slit is shown. 
Figure~\ref{Fig:EX2_StressStrain} presents the stress ($\sigma_{23}$, Left) and the strain ($\epsilon_{23}$, Right) values over the line $(0, 0.5)-(0.5,0.5)$  for different  $\beta$ values.
The stress and strain values are computed by Equation~\eqref{eq:airy:strcomp} and \eqref{str:nonlin}, respectively. 
We observe that the strain values are decreasing with increasing $\beta$-values, whereas the stress values increases close to the crack-tip. We emphasize that the growth of crack-tip strains, in NLSL, is not the same  order as the classical crack-tip square-root singularity. The, {Figure~\ref{Fig:EX2_StressStrain} delineates the expected crack-tip strain-limiting effect.

\begin{figure}[!h]
\centering
\includegraphics[width=1.0\textwidth]{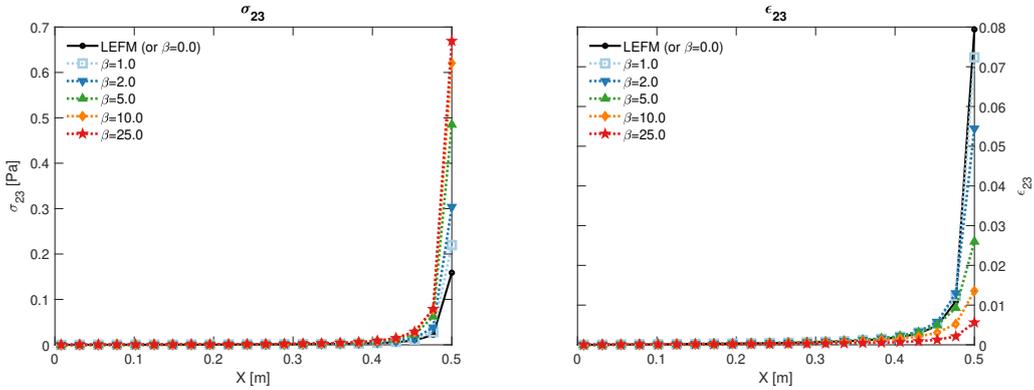}
\caption{Example 2. Stress ($\sigma_{23}$, Left) and strain ($\epsilon_{23}$, Right) on the center line in front of the static crack.}
\label{Fig:EX2_StressStrain}
\end{figure}

\subsection{Example 3: Static crack coupled with phase-field}
\begin{figure}[!h]
\centering
\begin{minipage}{0.45\textwidth}
\centering
\includegraphics[width=1.0\textwidth]{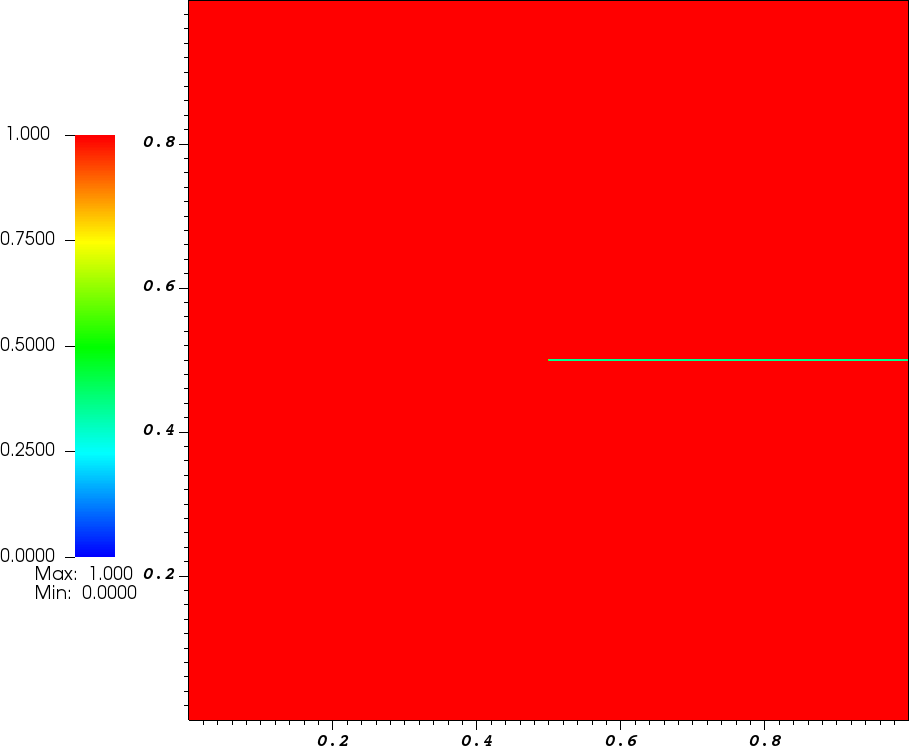}
\caption{Example 3. The phase-field crack presented over slit with $\varphi=0$.}
\label{Fig:Ex3_PF}
\end{minipage}
\hspace{1em}
\begin{minipage}{0.45\textwidth}
\centering
\includegraphics[width=0.85\textwidth]{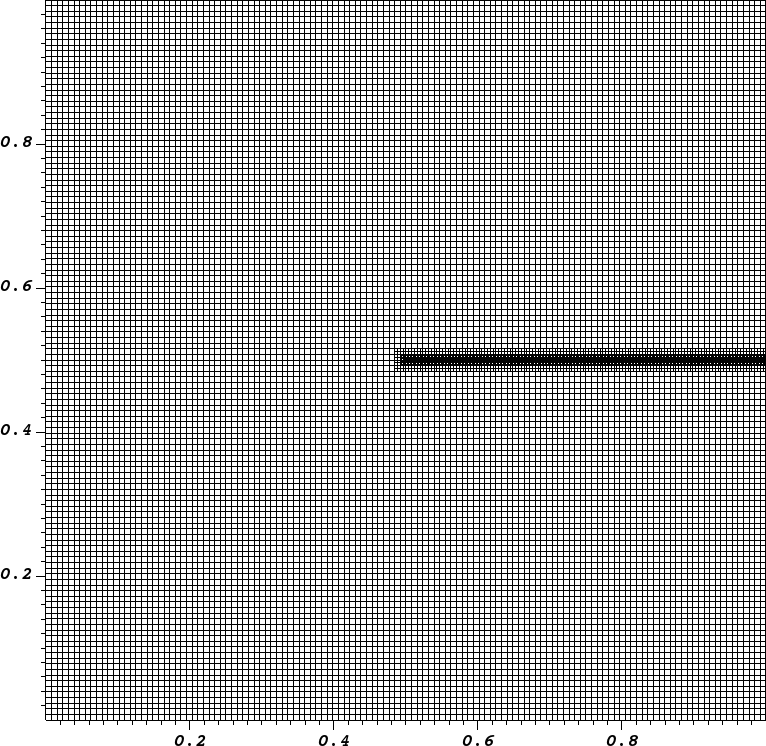}
\caption{Example 3. Adaptively refined mesh  around the crack.}
\label{Fig:Ex3_Mesh}
\end{minipage}
\end{figure}
\begin{figure}[!h]
\centering
\includegraphics[width=1.0\textwidth]{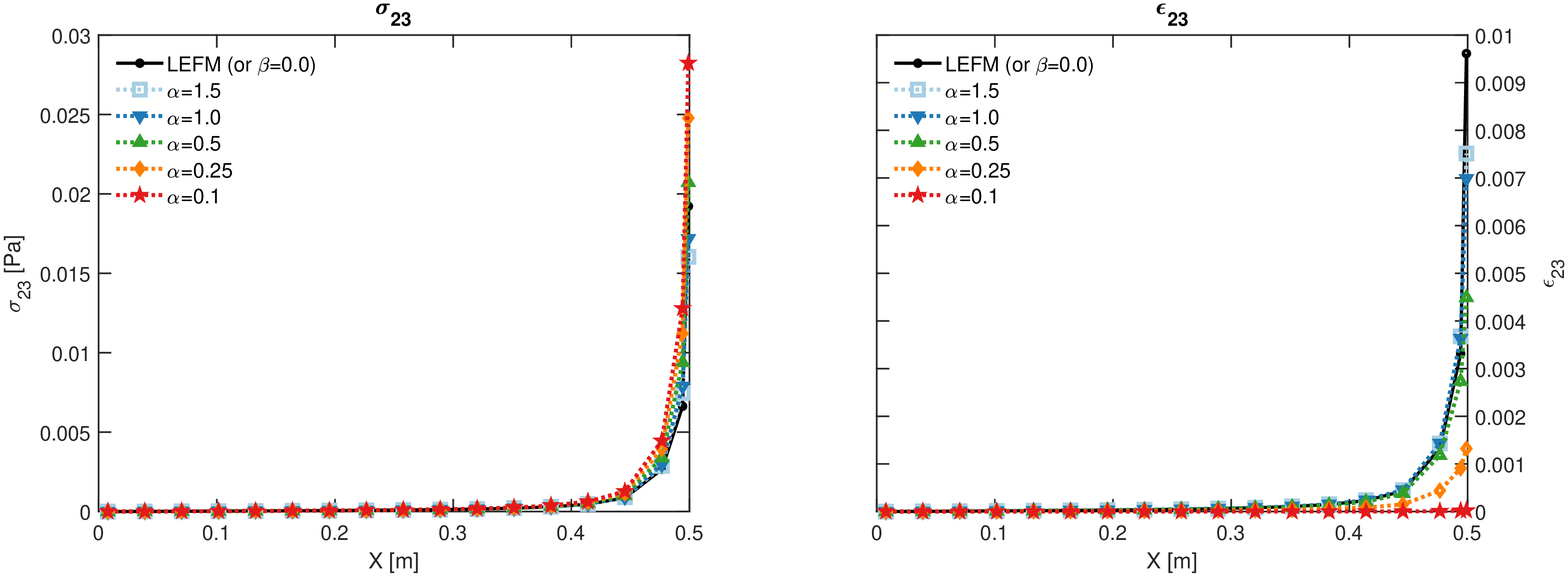}
\caption{Example 3. Stress ($\sigma_{23}$) and strain ($\epsilon_{23}$) on the center line for the static crack coupled with the phase-field.}
\label{Fig:EX3_StressStrain}
\end{figure}
In this example, we use a scalar valued phase-field variable to describe the fracture. 
In particular, we represent the initial slit in the previous example by a phase-field value, $\varphi = 0$. See Figure \ref{Fig:Ex3_PF}.
The level of global refinements is the same as previous example, and a small region containing the slit is additionally refined locally 
to perform a precise approximation (Figure \ref{Fig:Ex3_Mesh}). 
Here, the minimum cell diameter is given as $h_{min}= 0.00138107$,  the loading boundary conditions on the right-boundary is set as  $c=0.01$ {(see (Left) of Figure~\ref{Fig:ex2_setup})}, and the penalty parameter for the irreversibility of phase-field variable is given as  $\gamma=10^{4}$. 
We test a combination of nonlinear parameters where {$\beta=1.0$ and $\alpha=1.5, \, 1.0, \, 0.5, \, 0.25$ and $0.1$} 
and investigate the  behavior of stress-strain values near the crack-tip. 

With the given phase field values ($\varphi$), the computation of stress and strains are done using the following formulas:
\begin{subequations}\label{Eqs:Stress_Strain_PF}
\begin{align}
\sigma_{23} &= g(\varphi)  \Phi_{,1},   \\
\epsilon_{23} &= g(\varphi) \left( \frac{\Phi_{,1} }{2 \, \mu \left( 1 + \beta^{\alpha} \| \nabla \Phi \|^{\alpha} \right)^{1/\alpha}}    \right).
\end{align}
\end{subequations}

Although the phase-field regularization provides diffusive fracture, we still observe singular behavior of stress and strain values with the LEFM model ($\beta= 0$). However, with non-zero $\beta$ and $\alpha$-values, we observe that the growth of strain near the crack-tip is slower than the stress, which is the  distinctive feature of the model presented in this paper. 
{In addition, compared to Example 2 where we varied $\beta$-values to control the strain-limiting effects, 
we note that much slower growth is obtained with different $\alpha$-values.}

\subsection{Example 4: Quasi-static crack evolution}
Finally, we present a quasi-static crack propagation within the framework of the proposed nonlinear elasticity model. 
The computational domain and the mesh considered are depicted in the Figure~\ref{Fig:ex4_setup} and Figure~\ref{Fig:Ex4_Mesh}, respectively. 
\begin{figure}[!h]
\centering
\begin{minipage}{0.45\textwidth}
\centering
\begin{tikzpicture}

\node at (0.65, 3.85)   {$\Phi =ct$};
\node at (2.35, 3.85)   {$\Phi=-ct$};

\draw (0,0) -- (3,0) -- (3,3) -- (0,3) -- (0,0);
\draw [line width=0.5mm, blue]  (1.5,1.5) -- (1.5,3);
\draw [loosely dotted, line width = 0.5mm, red] (1.5,0) -- (1.5,1.5);
\node at (-0.3,-0.25)   {$(0,0)$};
\node at (3.3,3.2)   {$(1,1)$};
\node at (-0.3, 1.5)   {$\Lambda_L$};
\node at (3.3, 1.5)   {$\Lambda_R$};
\node at (1.5, -0.25)   {$\Lambda_{B}$};
\node at (1.5, 3.2)   {$\Lambda_{T}$};
\node at (1.9, 2.0)   {$\Gamma_{C}$};
\node at (2.0, 3.3) {$\bigotimes$};
\node at (2.7, 3.3) {$\bigotimes$};
\node at (0.3, 3.3) {$\bigodot$};
\node at (1.0, 3.3) {$\bigodot$};
\end{tikzpicture}
\caption{Example 4. A setup and the boundary conditions are illustrated. The bold blue line denotes the initial fracture described by the slit and the red-dotted line is the expected crack path. Tractions for the anti-plane shear are on the $\Lambda_{T}$.}
\label{Fig:ex4_setup}
\end{minipage}
\hspace{1em}
\begin{minipage}{0.45\textwidth}
\centering
\includegraphics[width=0.8\textwidth]{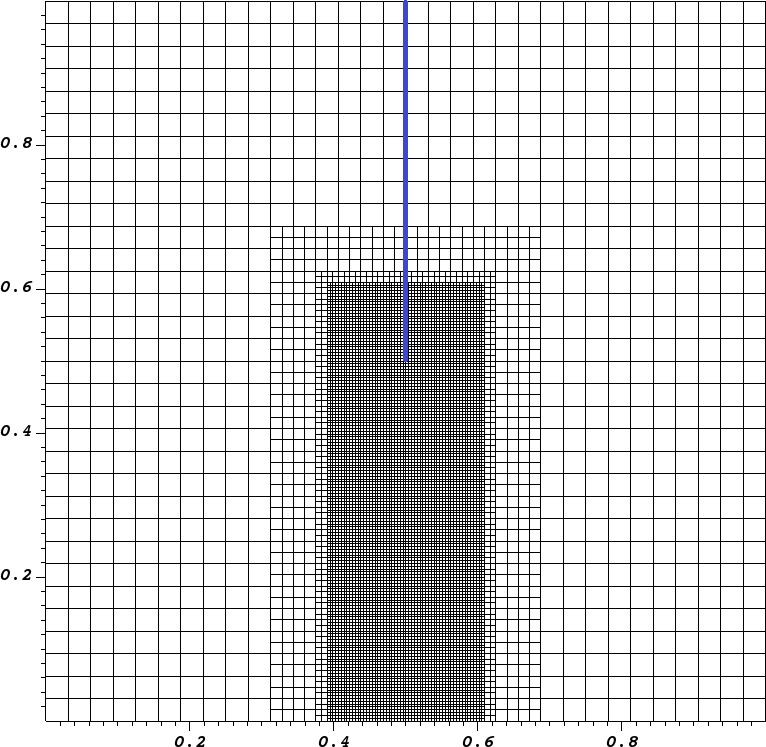}
\caption{Example 4.  Adaptively refined mesh around the crack and the crack path.}
\label{Fig:Ex4_Mesh}
\end{minipage}
\end{figure}
The physical and numerical parameters are set as shown in the following Table~\ref{Tab:Exs-4}. }
\begin{table}[!h]
\begin{center}
\small
\caption{{Parameters for Example 4.}}
\label{Tab:Exs-4}
\begin{tabular}{l|clc}
\toprule
\textbf{Parameter} & \textbf{Value} & \textbf{Unit}\\
\midrule
Shear modulus ($\mu$) & 20.0 & $Pa$\\
Critical energy release rate ($G_c$) & 1.0 & $Nm^{-1}$\\
Gamma for the penalty ($\gamma$) & 1.0e-4 & - \\
Traction coefficient  ($c$)  &25.0   & - \\
Time step size ($\Delta t$) & 1.0e-2 & - \\
\bottomrule
\end{tabular}
\end{center}
\end{table}

The quasi-static loading conditions on top boundary are set as $\Phi = ct$, where the magnitude of loading increases in time with $\Delta t$.
With the adaptive mesh refinement,  the smallest mesh is given as  $h_{min}= 0.00552427$. 
In this example, we test four different combinations of the nonlinear parameter values of ($\alpha$, $\beta$): 
case i) ({$\alpha \neq 0$, $\beta=0$}),
{case ii)} ($\alpha = 0.5$,  $\beta=0.001$),
{case iii)} ($\alpha = 0.5$, $\beta=0.003$), and
{case iv)} ($\alpha = 0.3$,  $\beta=0.001$). 
{Thus, case i) represents LEFM model, and case ii)-iv) are NLSL models. }

First, Figure \ref{fig:ex4_propagations} presents the propagating fractures for each case at given time steps. 
The (Top Row) presents the phase-field fracture for the case i), LEFM, and the other rows are for each of the NLSL cases ii) - iv).  
We note that the fracture initiation time and the speed of propagation vary based on the nonlinear parameters. 
{In addition, Figure \ref{fig:ex4_quasi_LEFM_NLSL} presents the comparison of the phase-field and strain values for each cases at a fixed time given as $t=0.32$.}
We confirm that LEFM (case i) has the earliest initiation of crack, whereas NLSL with case iv) $(\alpha,\beta)=(0.3,0.001)$ has the latest. 
More details of the crack-tip discrete speed  is compared in Figure \ref{Fig:EX4_Energy_Speed} (Left).

\begin{figure}[!h]
\centering
\subfloat[case i) $n=32$]{\includegraphics[width= 0.35\textwidth]{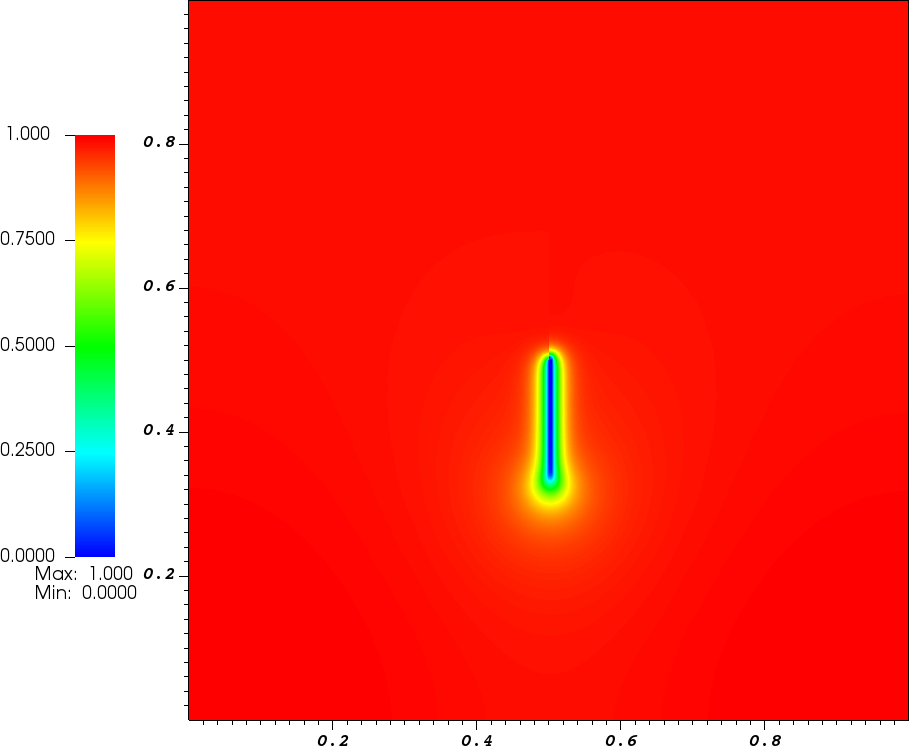}}
\hspace{0.1in}
\subfloat[case i)  $n=33$]{\includegraphics[width= 0.35\textwidth]{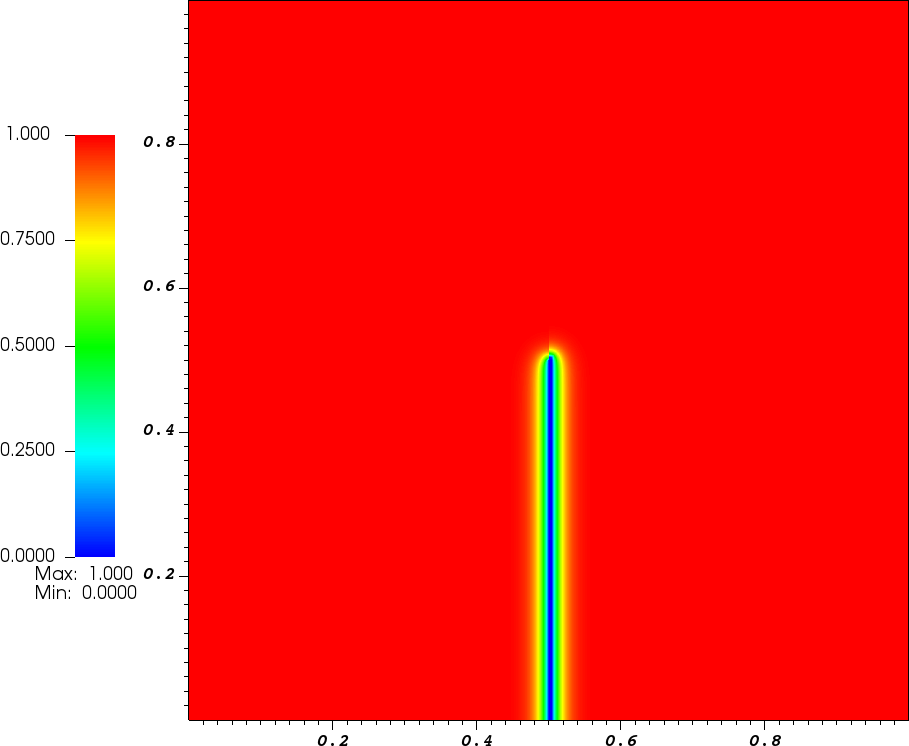}}\\
\subfloat[case ii)  $n=37$]{\includegraphics[width= 0.35\textwidth]{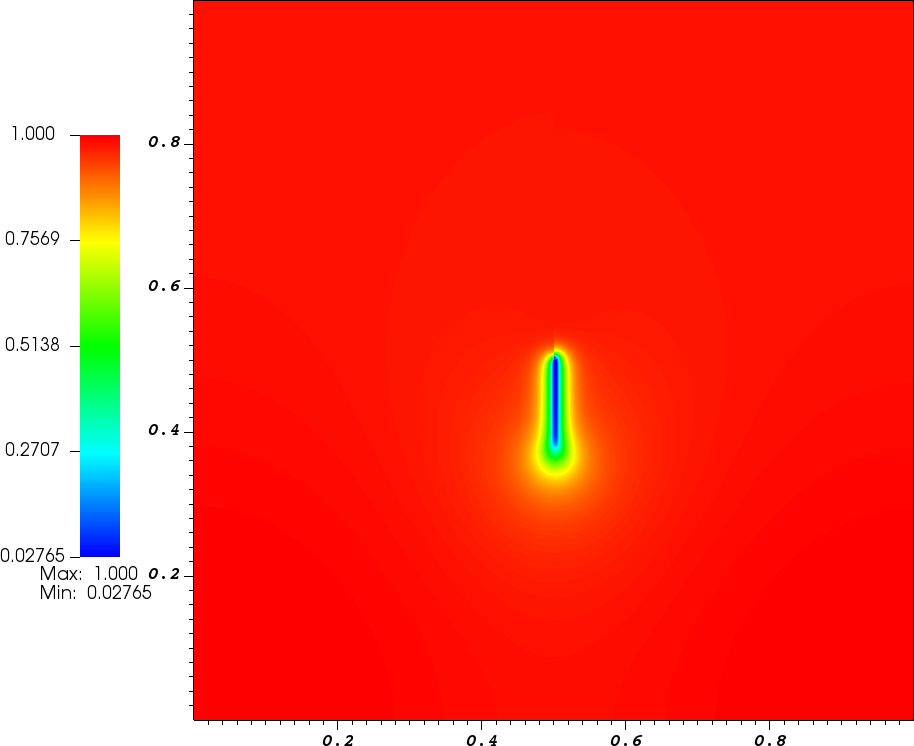}}
\hspace{0.1in}
\subfloat[case ii)  $n=38$]{\includegraphics[width= 0.35\textwidth]{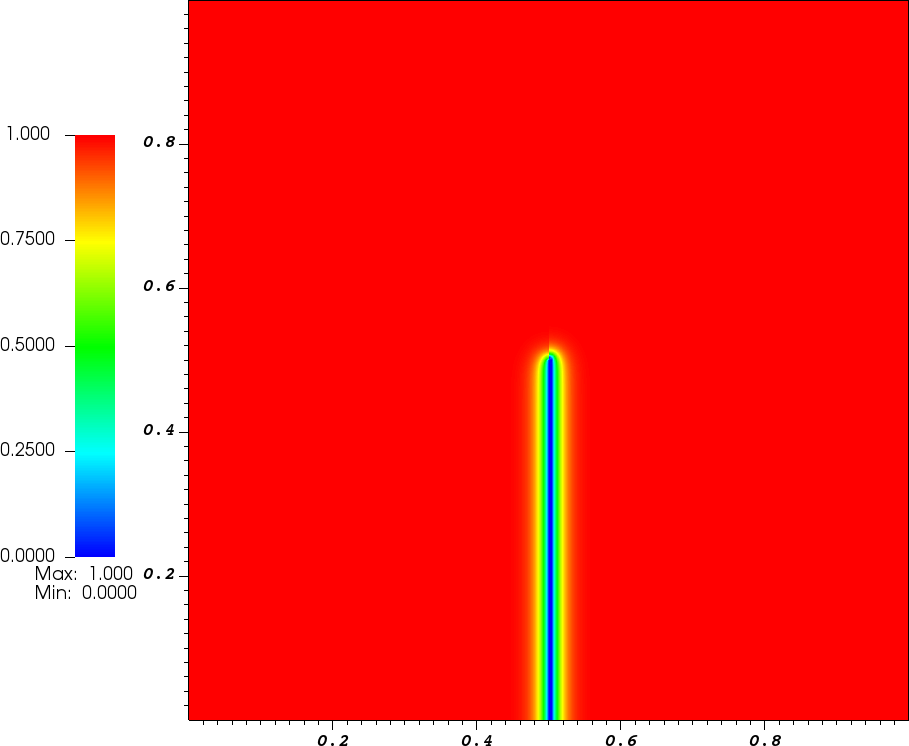}} \\
\subfloat[case iii)  $n=41$]{\includegraphics[width= 0.35\textwidth]{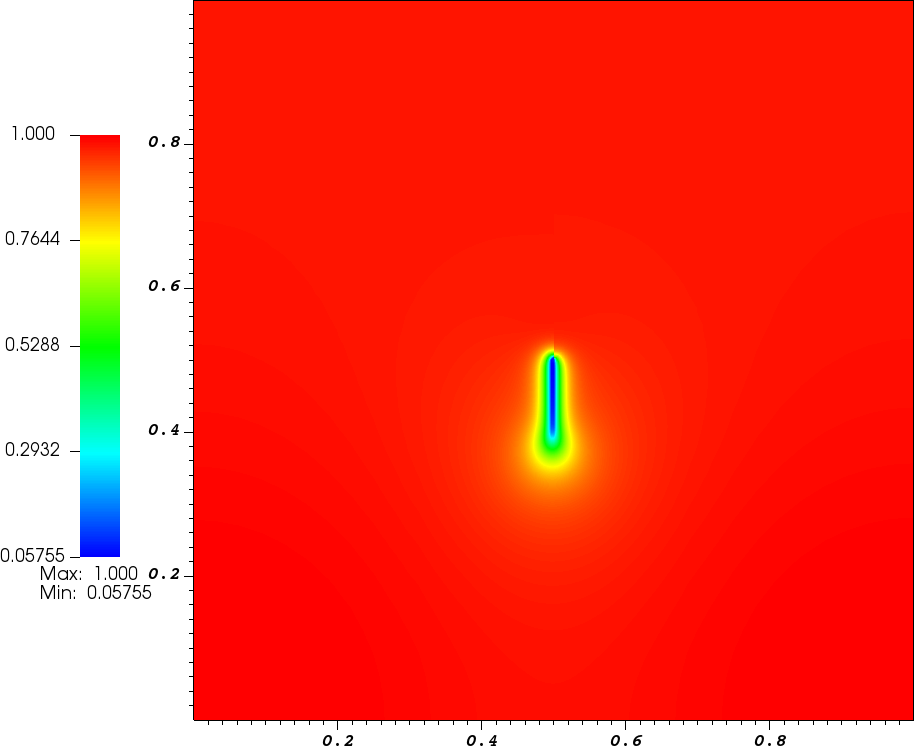}}
\hspace{0.1in}
\subfloat[case iii)  $n=43$]{\includegraphics[width= 0.35\textwidth]{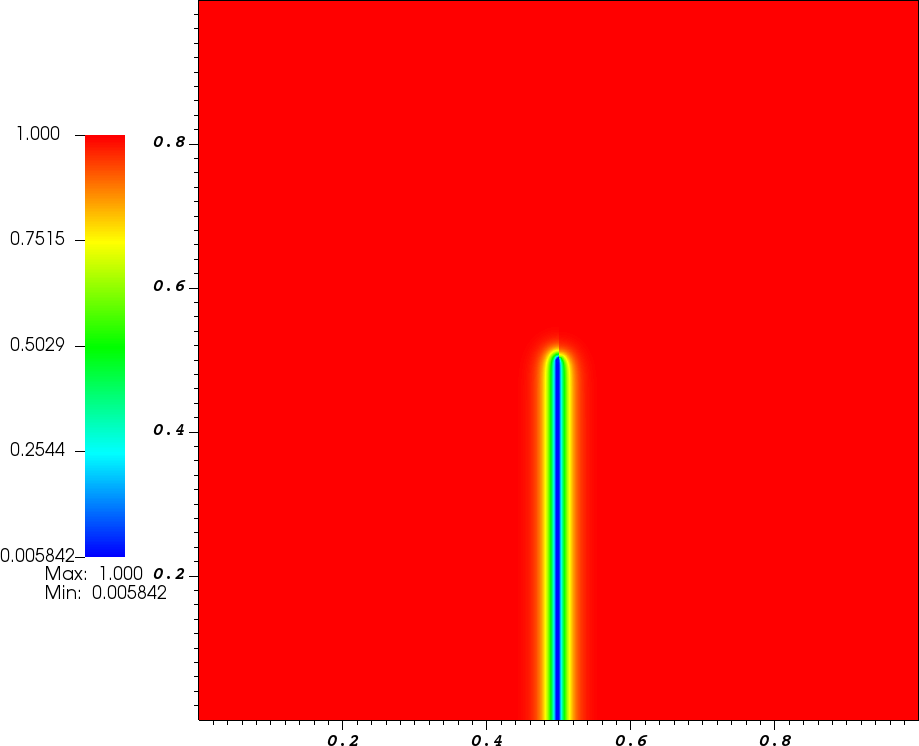}}\\
\subfloat[case iv)  $n=53$]{\hspace{0.5em}\includegraphics[width= 0.35\textwidth]{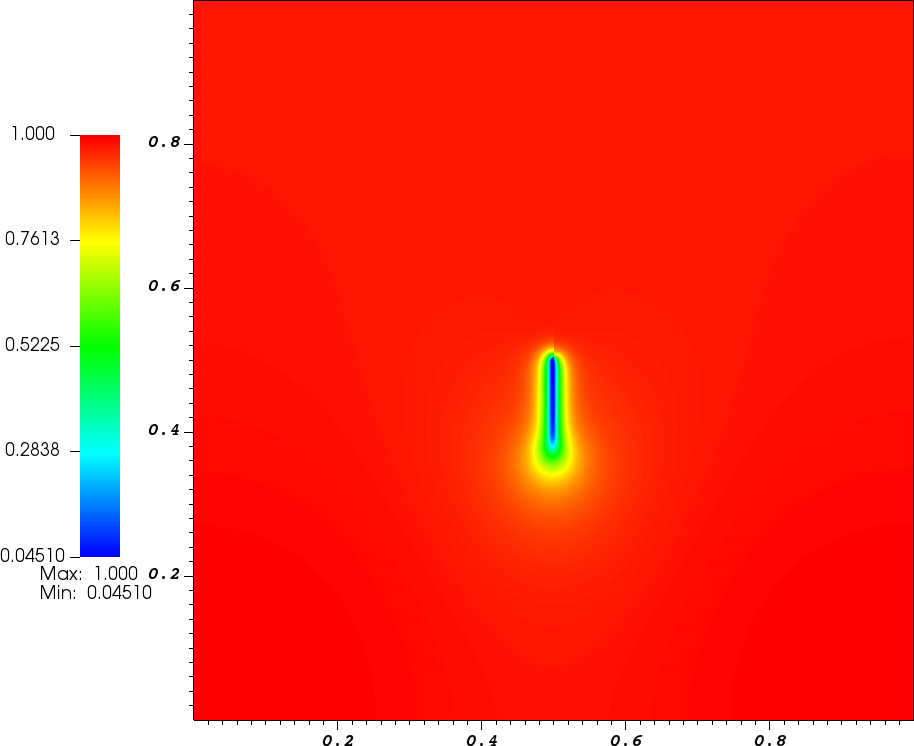}}
\hspace{0.1in}
\subfloat[case iv)  $n=54$]{\includegraphics[width= 0.35\textwidth]{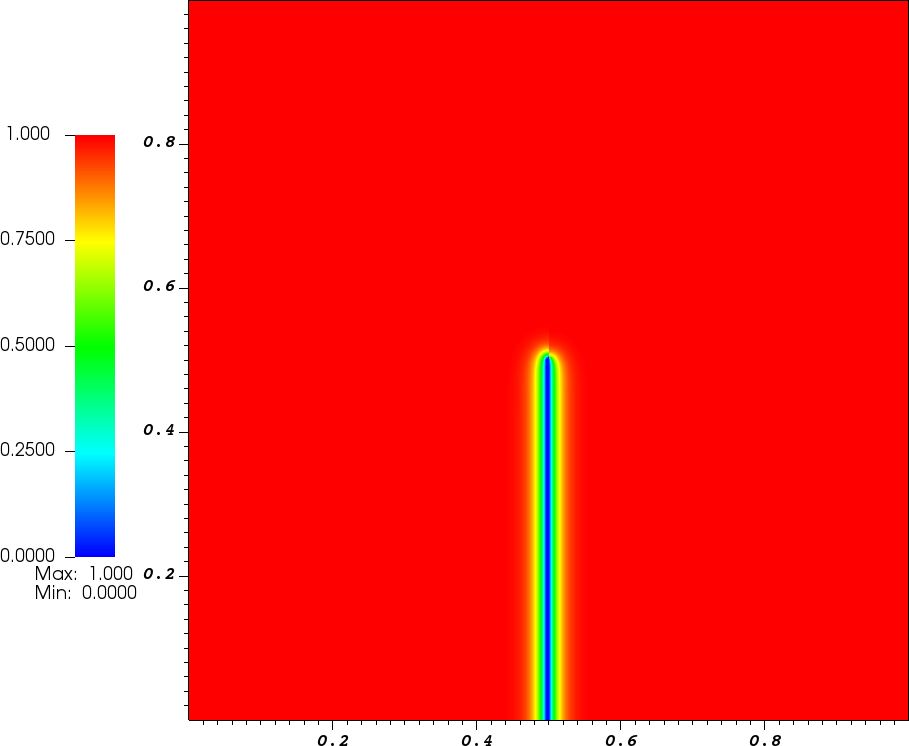}}
\hspace{0.1in}
\caption{Example 4. Phase-field values of propagating fractures for each case: 
(Top Row) case i), (Second Row) case ii), (Third Row) case iii), and (Bottom Row) case iv). The fracture initiation time 
varies based on 
the nonlinear parameters. }
\label{fig:ex4_propagations}
\end{figure}

\clearpage
\newpage

\begin{figure}[!h]
\centering
\subfloat[case i) $\varphi$]
{\includegraphics[width= 0.35\textwidth]{linpf32}}
\hspace*{0.2in}
\subfloat[case i) $\epsilon_{23}$]
{\includegraphics[width=0.35\textwidth]{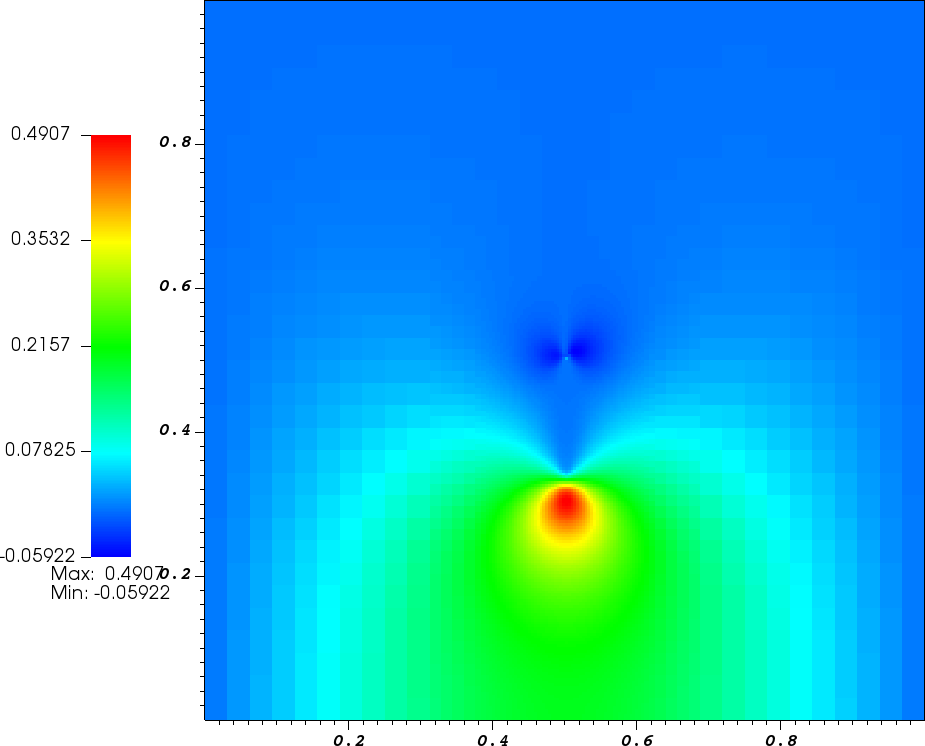}} \\
\subfloat[case ii) 
$\varphi$]
{\includegraphics[width= 0.35\textwidth]{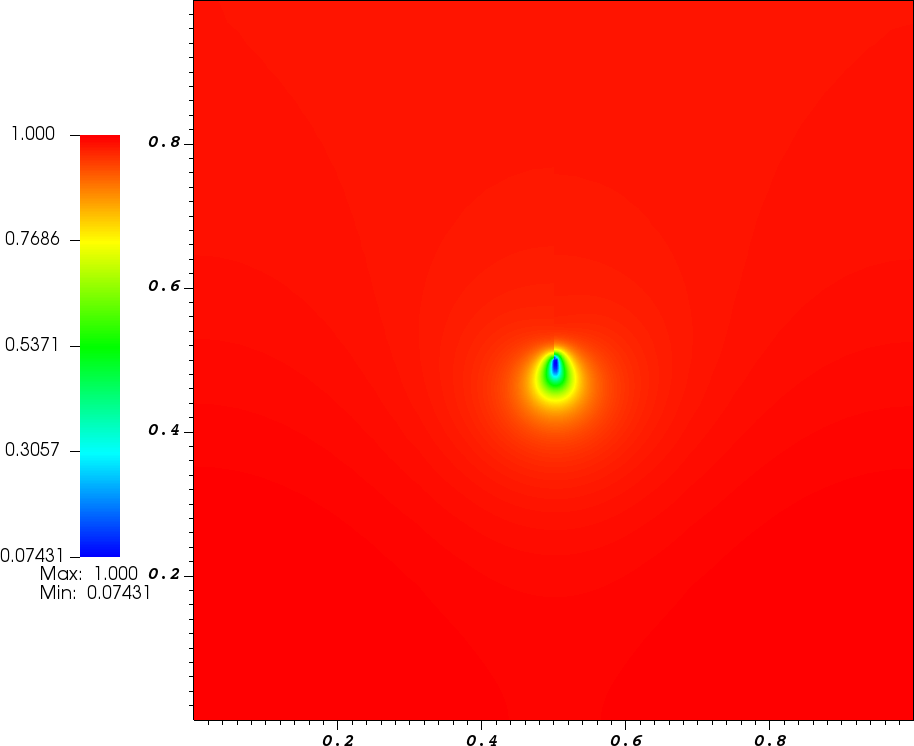}}
\hspace*{0.2in}
\subfloat[case ii) 
$\epsilon_{23}$]
{\includegraphics[width=0.35\textwidth]{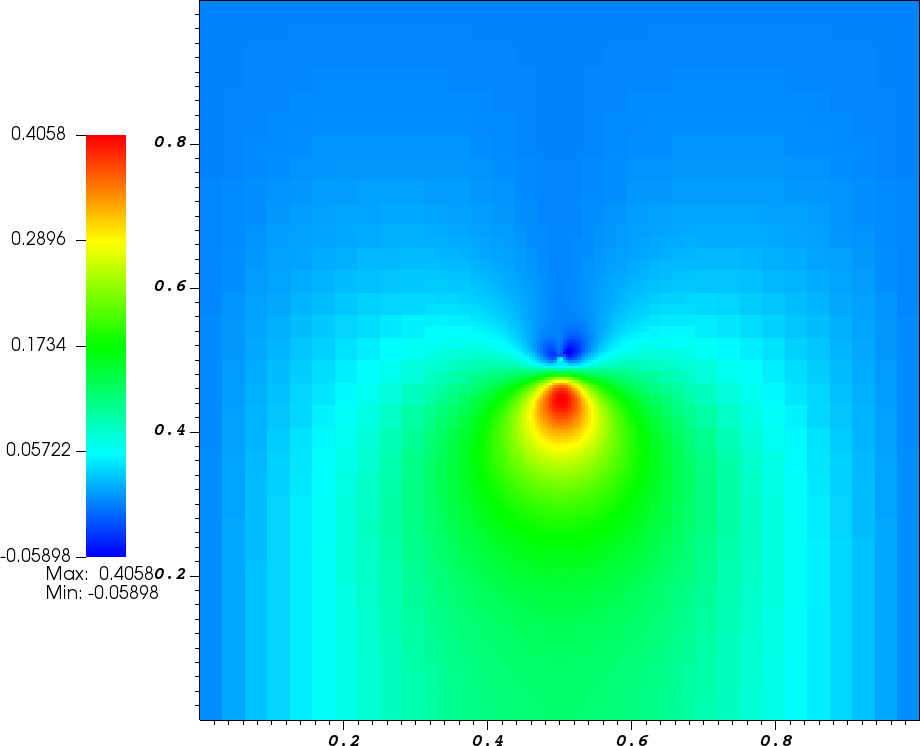}}  \\
\subfloat[case iii) 
$\varphi$]
{\includegraphics[width= 0.35\textwidth]{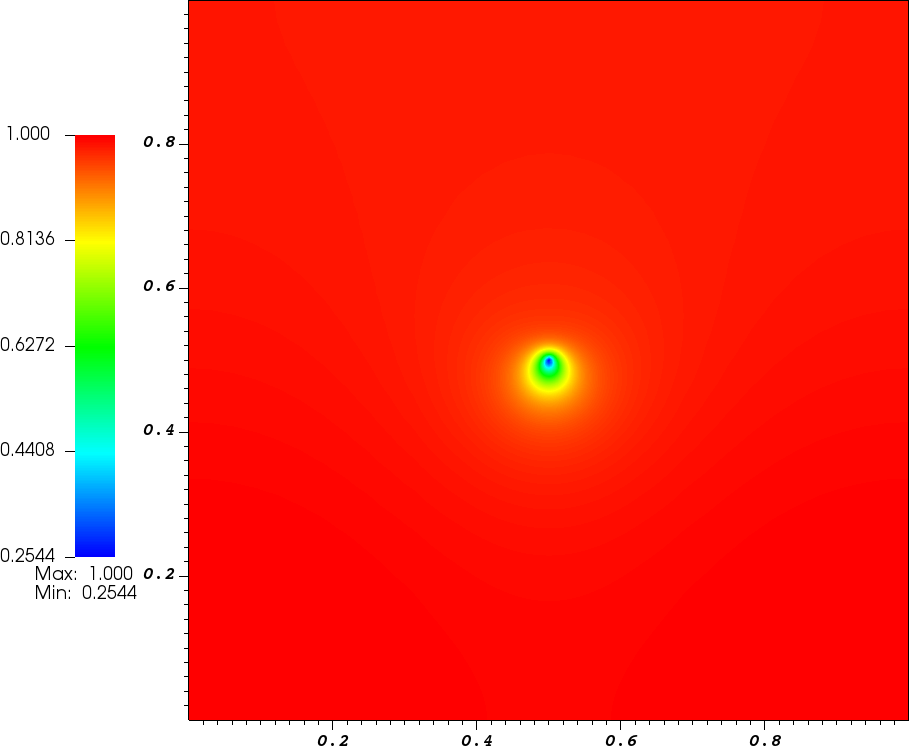}}
\hspace*{0.2in}
\subfloat[case iii) 
$\epsilon_{23}$]
{\includegraphics[width=0.35\textwidth]{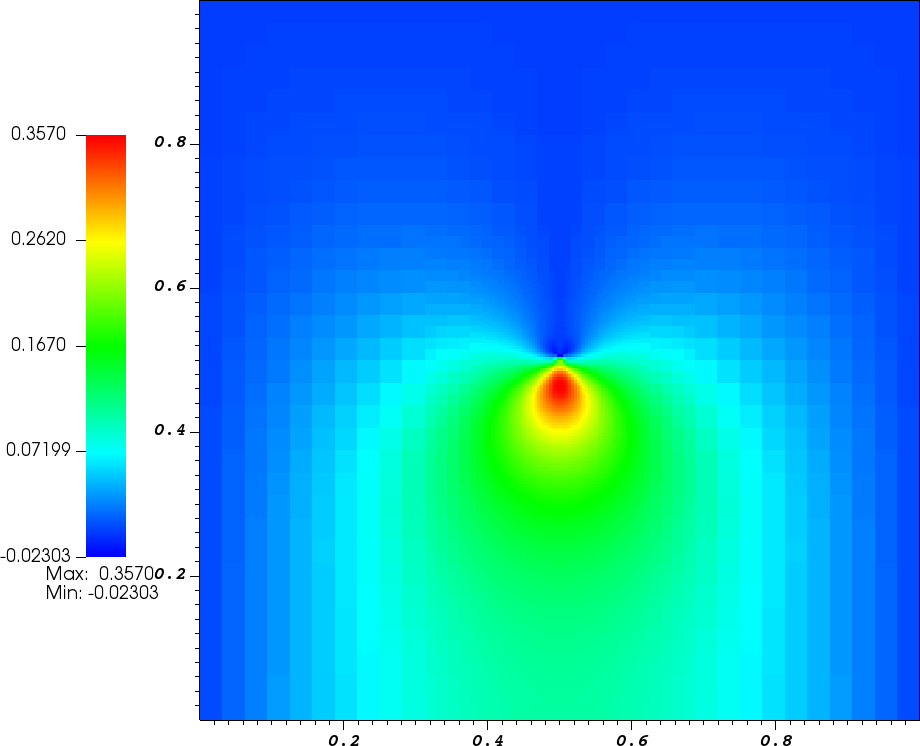}} \\
\subfloat[case iv) 
$\varphi$]
{\includegraphics[width= 0.35\textwidth]{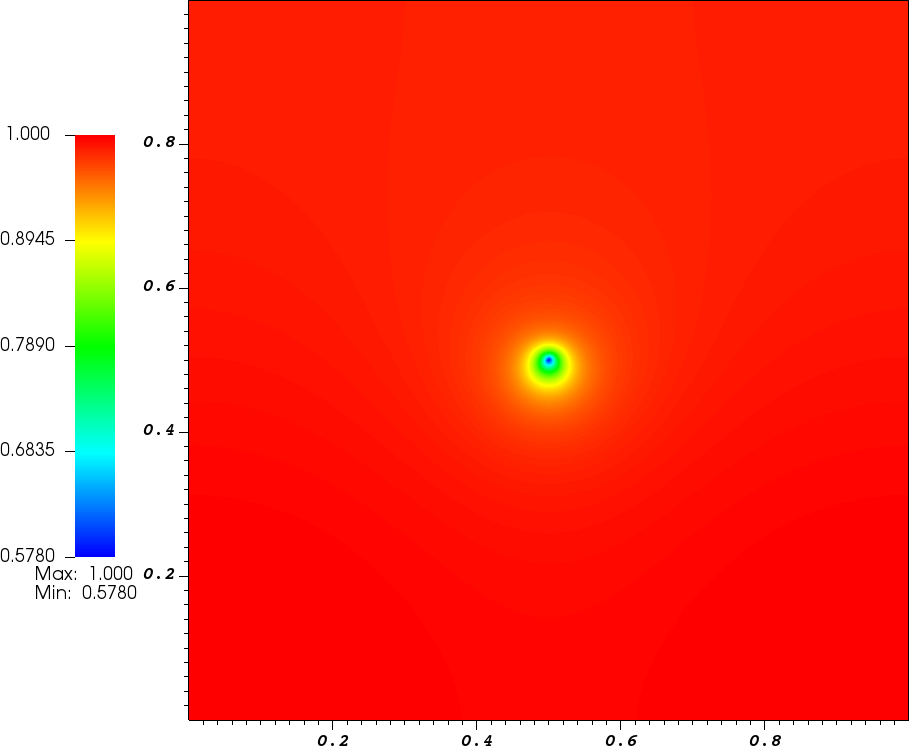}}
\hspace*{0.2in}
\subfloat[case iv) 
$\epsilon_{23}$]
{\includegraphics[width=0.35\textwidth]{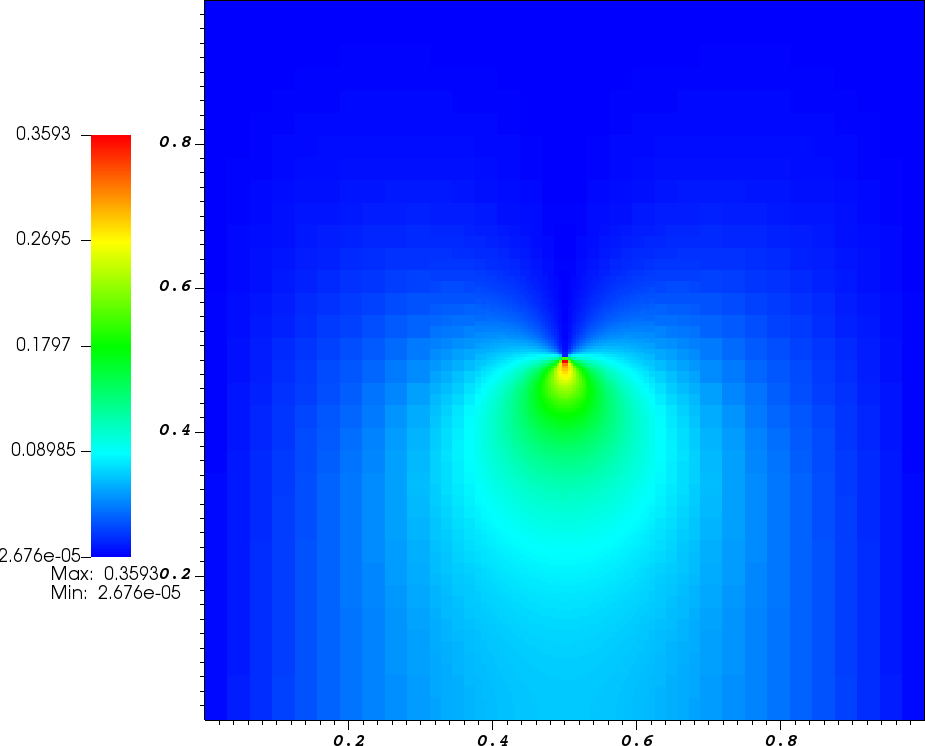}} 
\caption{Example 4. Snapshots for each case at the given time $t=0.32$ ($n=32$). 
(Left Column) illustrates the phase-field values during crack evolution.
(Right Column) presents the corresponding $\epsilon_{23}$ values for each case.
}
\label{fig:ex4_quasi_LEFM_NLSL}
\end{figure}

\clearpage
\newpage

\begin{figure}[!h]
\centering
\includegraphics[width=0.45\textwidth]{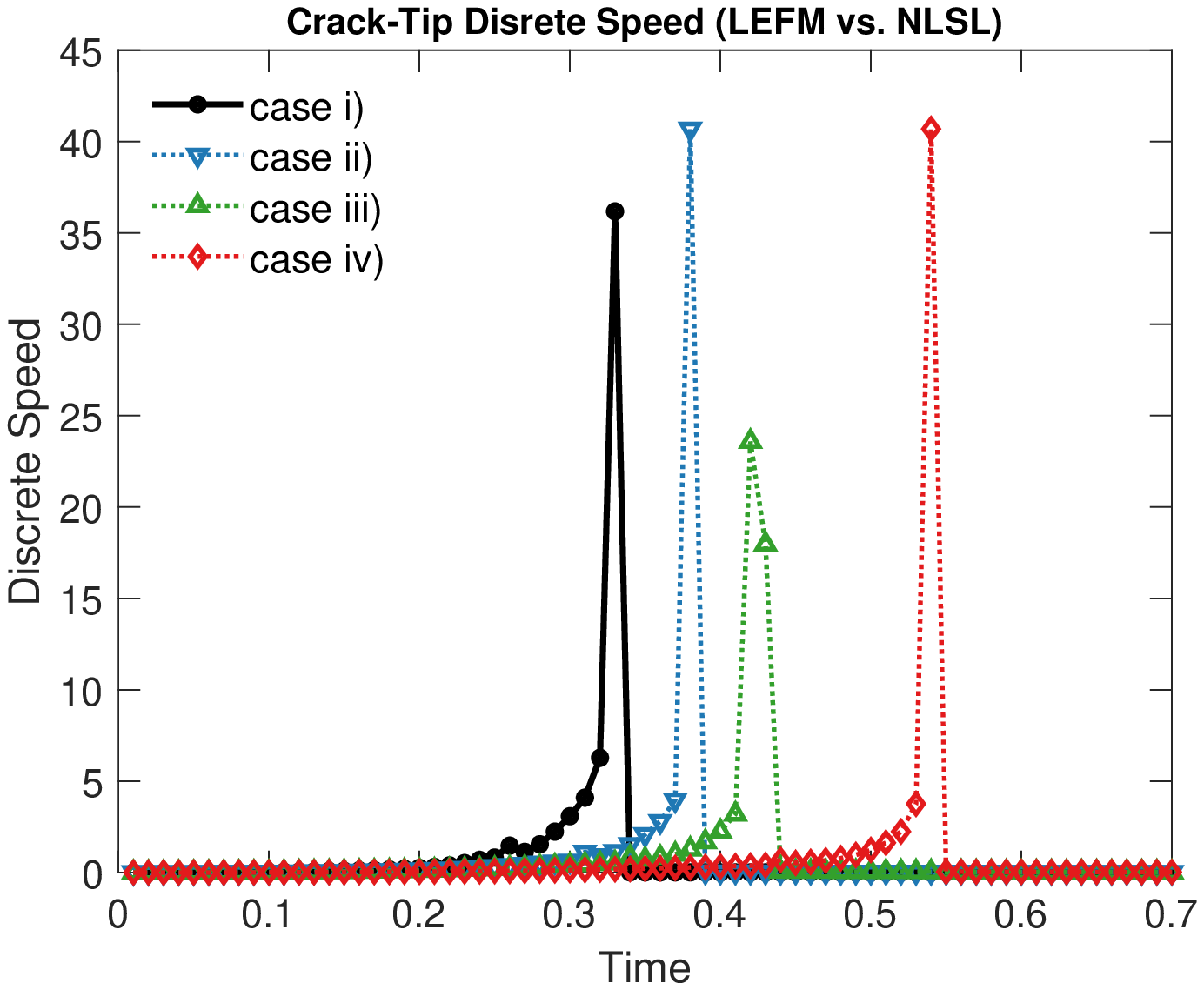}
\includegraphics[width=0.45\textwidth]{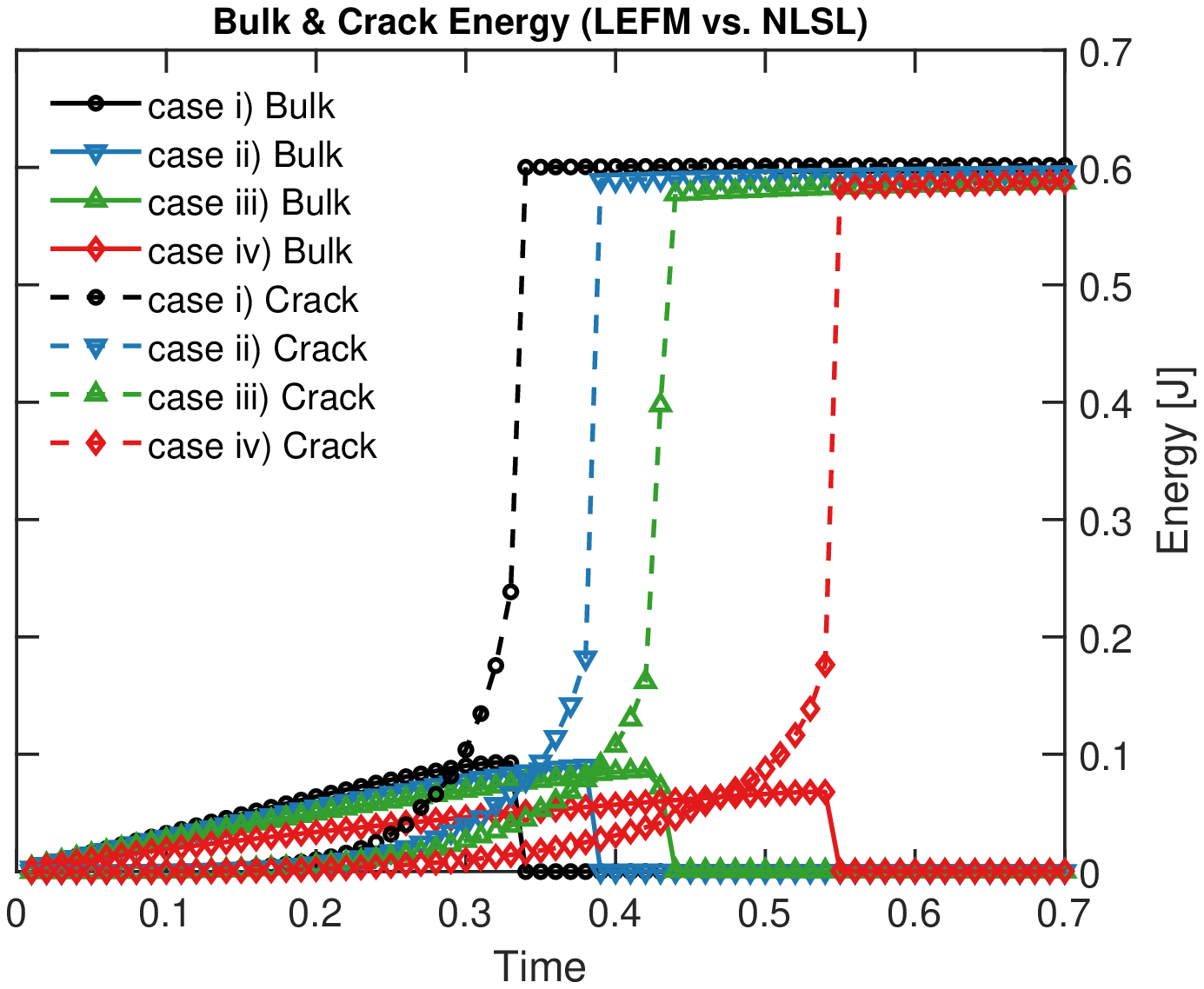}
\caption{Example 4. (Left) discrete crack speed and (Right) bulk and crack energies for each case are illustrated.}
\label{Fig:EX4_Energy_Speed}
\end{figure}

Next, we emphasize that the different initiation of fracture and its propagation speed are due to the distinct energy balance between bulk (non-crack) and crack energies for each case.  
Based on Equation~\eqref{reg:energy}, the nonlinear bulk energy for NLSL is defined as
\begin{equation}\label{eqn:NLSL_bulk_energy}
\frac{1}{2} \int_{\Omega} \left( (1- \kappa) \varphi^2 + \kappa  \right) \; \frac{ \| \nabla \Phi^n \|^2}{2 \, \mu \left( 1 + \beta^\alpha \; \| \nabla \Phi^n \|^{\alpha} \; \right)^{1/\alpha}} \; d \bfx,
\end{equation}
whereas the crack energy is defined as
\begin{equation}\label{eqn:NLSL_crack_energy}
 G_c \int_{\Omega} \left[ \frac{(1-\varphi)^2}{2\xi} + \frac{\xi}{2} \; |\nabla \varphi|^2  \right]  \; d\bfx.
\end{equation}
Equation~\eqref{eqn:NLSL_bulk_energy} shows that the 
bulk energy will increase slower with more strain-limiting effects with lager $\beta$-values or smaller $\alpha$-values. 
Figure~\ref{Fig:EX4_Energy_Speed} (Right) presents the computed bulk and the crack energies for each case along the simulation time. 
{We observe that the two energies from the NLSL models are slightly smaller than the LEFM,
which is related to smaller displacements and strains from the strain-limiting effects for NLSL. }

\clearpage
\newpage

\begin{figure}[!h]
\centering
\includegraphics[width=0.95\textwidth]{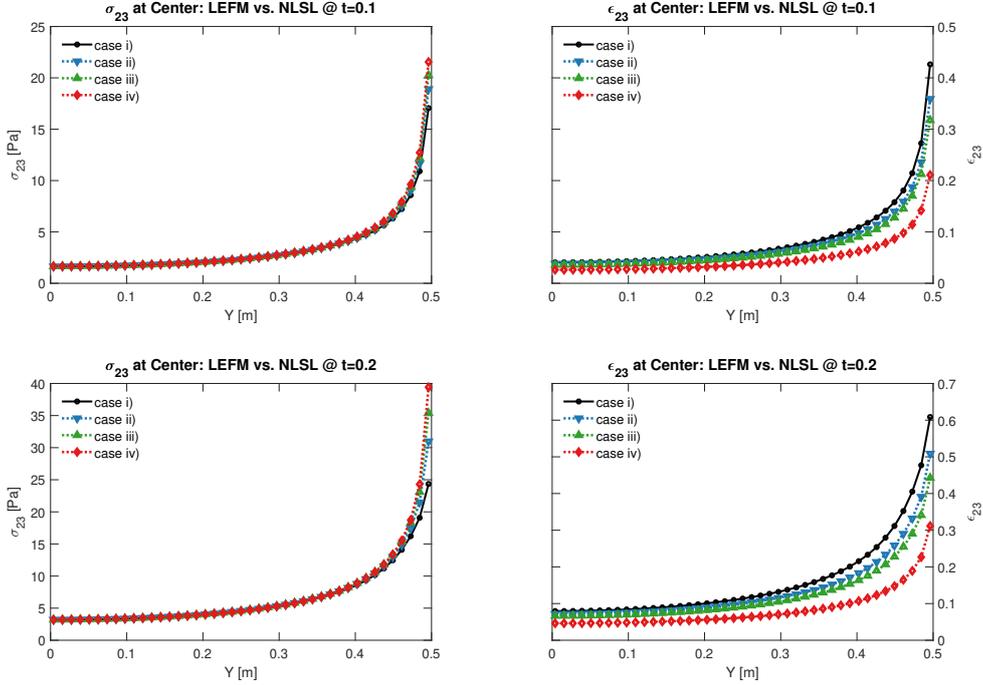}
\caption{Example 4. Stress ($\sigma_{23}$, Left Column) and strain ($\epsilon_{23}$, Right Column) on the center line for t=0.1 (Top Row) and  t=0.2 (Bottom Row).}
\label{Fig:EX4_CenterStressStrain_10_20}
\end{figure}

Next, Figure~\ref{Fig:EX4_CenterStressStrain_10_20} illustrates the comparison of stress (Left Column) and strain (Right Column) for each case.  The stress and strain values are computed along the mid-line (red-dotted line in Figure~\ref{Fig:ex4_setup}) directly ahead of the crack-tip, 
and the values correspond to the time step $n=10$ for the (Top Row) and the time step $n=20$ for the (Bottom Row).  
For the process of propagating fractures, we still observe that the strain values for NLSL are more limited compared to LEFM and do not increase in the same order as the stress values at the crack-tip. 

{For an in-depth study, stress and strain curves at a fixed location as the original crack-tip are plotted in Figure~\ref{Fig:EX4_CrackTipStressStrain} along the simulation time (each data point indicates corresponding time step). Recall that the stress and strain values are computed by Equations \eqref{Eqs:Stress_Strain_PF} where $g(\varphi)$ is defined as Equation \eqref{eqn:g_phi}. 
While the fracture propagates, we note that the phase-field value decreases at the crack-tip, (i.e., $\varphi=1\rightarrow\varphi =0$), thus the magnitude of $g(\varphi)$ also decreases. The blue circle on the right side of each plot for each case indicates when the case reaches the maximum stress value, and the red circle on the left indicates the time step right before it plunges abruptly (stress drop from propagation). Both are denoted with their specific time step numbers for each case.}

}

\begin{figure}[!h]
\centering
\includegraphics[width=0.65\textwidth]{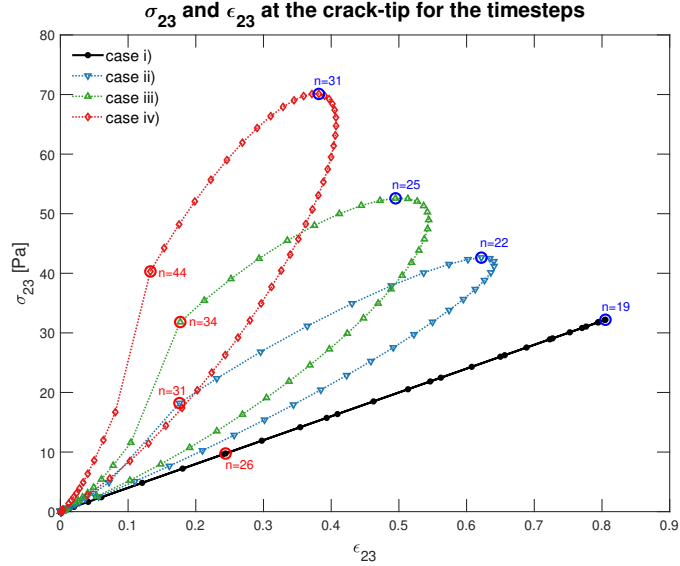}
\caption{Example 4. Stress ($\sigma_{23}$) and strain ($\epsilon_{23}$) curve for each case at the crack-tip for the time steps when the phase-field ($\varphi)$ value drops from $1$ to $0$. Blue circle denotes the time step for its maximum stress value, and red circle indicates the time step before the stress drop for each case. 
}
\label{Fig:EX4_CrackTipStressStrain}
\end{figure}

{First, we find that NLSLs have different stress-strain curves than that of LEFM in their shapes. Each NLSL has some intent of hysteresis in stress and strain due to its nonlinearity, while stress and strain values of LEFM stay on the same line over the time steps. Particularly for NLSL, we also note that the strain value starts to decrease before the stress reaches its maximum (blue circle), and this is due to the decreased $g(\varphi)$ in $\epsilon_{23}$ for small strain value growing with much smaller order compared to stress increasing with the singular behavior. Further, we note that the initiations of stress drop due to the initiations of crack growths are clearly illustrated in the NLSL curves (see after red circles), whereas no such distinctive one is for LEFM. (For LEFM, it can be found through $\sigma_{23}$ values against time steps, which can be more clearly represented through the values of $\sigma_{23}$ divided by $g(\varphi)$.) Thus, between the blue and red circles, the relatively smooth decreasing of stress value in each case of NLSL is due to the decreased $g(\varphi)$ in $\sigma_{23}$. {Again, the most strain-limiting effect with slower initiation of fracture can be seen 
for case iii) with $(\alpha,\beta)=(0.3, 0.001)$, where the difference for the maximum strain value at the original crack-tip is 
about $0.4$ 
between LEFM and NLSL (Figure~\ref{Fig:EX4_CrackTipStressStrain}). 
}}

\clearpage
\newpage

\section{Conclusion}
\label{sec:conclusion}

A major covet of this paper is to investigate physical models for the 
evolution of static crack in the brittle elastic materials. 
Towards that end, recently introduced strain-limiting models based on the implicit theories offer an attractive feature, in that the strains are uniformly bounded in the body. Moreover, a main goal of this work is to integrate the energy minimization-based phase-field regularization with the bulk energy being modeled by the nonlinear 
elasticity. 
For coupling nonlinear mechanics with phase-field, we utilize an iterative staggered method, i.e., the L-scheme, and an augmented Lagrangian method to accommodate the crack-irreversibility. 
Our 
numerical experiments 
demonstrate the strain-limiting effects with much slower growth than the stress near the crack-tip as 
expected by the model. In addition, compared to the classical LEFM, we 
observe that fracture propagation speed and the initiation of the crack of the proposed model 
can depend on the nonlinear modeling parameters. 
More detailed investigation for the effect of nonlinear modeling parameters and comparison with the experiments are part of ongoing works. 

\section*{Acknowledgements}
\label{sec:acknowledgements}
This research done by S. Lee is based upon work supported by the National Science
Foundation under Grant No. (NSF DMS-1913016). Other authors, Hyun C. Yoon and S.
M. Mallikarjunaiah, would like to thank the support of College of Science \& Engineering,
Texas A\&M University-Corpus Christi.

\bibliography{airynonlinear}
\end{document}